\documentclass[aps, pre, twocolumn, a4paper, superscriptaddress, floatfix]{revtex4}
\usepackage{graphicx}
\usepackage{amsmath}
\usepackage{placeins}
\usepackage{color}
\usepackage{hyperref}

\newcommand{\er}{Erd\H{o}s-R\'{e}nyi}
\newcommand{\kk}{\bar k }

\newcommand{\scol}{S_{\rm color}}
\newcommand{\lcol}{\mathcal{L}_{\rm color}}

\newcommand{\beq}{\begin{equation}}
\newcommand{\eeq}{\end{equation}}
\newcommand{\beqa}{\begin{eqnarray}}
\newcommand{\eeqa}{\end{eqnarray}}

\begin{document}
\title{Optimal redundancy against disjoint vulnerabilities in networks}
\author{Sebastian M. Krause}
\affiliation{Theoretical Physics Division, Rudjer Bo\v{s}kovi\'{c} Institute, Zagreb, Croatia}
\author{Michael M. Danziger}
\affiliation{Department of Physics, Bar Ilan University, Ramat Gan, Israel}
\author{Vinko Zlati\'{c}}
\affiliation{Theoretical Physics Division, Rudjer Bo\v{s}kovi\'{c} Institute, Zagreb, Croatia}


\begin{abstract}
Redundancy is commonly used to guarantee continued functionality in networked systems.
However, often many nodes are vulnerable to the same failure or adversary.
A ``backup'' path is not sufficient if both paths depend on nodes which share a vulnerability.
For example, if two nodes of the Internet cannot be connected without using routers belonging to a given untrusted entity, then all of their communication--regardless of the specific paths utilized--will be intercepted by the controlling entity.
In this and many other cases, the vulnerabilities affecting the network are disjoint:
each node has exactly one vulnerability but the same vulnerability can affect many nodes. 
To discover optimal redundancy in this scenario, we describe each vulnerability as a color and develop a ``color-avoiding percolation'' which uncovers a hidden color-avoiding connectivity.
We present algorithms for color-avoiding percolation of general networks and an analytic theory for random graphs with uniformly distributed colors including critical phenomena.
We demonstrate our theory by uncovering the hidden color-avoiding connectivity of the Internet.
We find that less well-connected countries are more likely able to communicate securely through optimally redundant paths than highly connected countries like the US.
Our results reveal a new layer of hidden structure in complex systems and can enhance security and robustness through optimal redundancy in a wide range of systems including biological, economic and communications networks.
\end{abstract}
\maketitle

\section{Introduction}
Many real-world complex systems, which we model as networks, display disjoint vulnerability to failure or attack.
These vulnerabilities make networks far less robust than they seem.
It is generally assumed that redundant connections through multiple paths improves robustness \cite{stelling-cell2004,carmi-pnas2007} but if a given vulnerability affects a large set of nodes, this may not be the case.
For example, if one node can not communicate with another without routing the information through routers under a given entity's control, secure communication is compromised.
Similarly, in an economic network, if a firm has redundant suppliers but each supply chain includes nodes belonging to a given company, then there is an absence of competition--even if in principle there are multiple competing companies working in that sector.
Similar considerations hold for nodes in a spatial network that are located near one another  because 
transportation and economic assets in the same city will be affected by the same weather events or disasters \cite{neumayer-milcom2008,agarwal-infocom2011,berezin-scireps2015}. 
Disjoint vulnerabilities also appear in  biological networks. 
Depending on the type of nutrients available, different metabolic pathways are enabled \cite{Schuster2000Metabolic,Feil2012epigenetics}. 
In this case, the metabolic network is disjointly vulnerable to the absence of a certain type of nutrient. 
Robust functionality can be guaranteed only if there are paths  connecting source and target metabolite even when each distinct nutrient is removed. Gene regulatory networks exhibit similar multipath responses to environmental conditions \cite{pal-nature2006,white-cell2013}.


In all of these cases, connectivity alone gives a poor picture of the network's robustness and security.
However, since susceptibility to one vulnerability often precludes susceptibility to another vulnerability,  we can partition the network into disjoint subsets by vulnerability.
The disjoint nature of the vulnerabilities allows for robust connectivity to be established, provided the network remains connected when each subset is removed.
Here we present a new framework for analyzing disjointly vulnerable complex networks and show the conditions for which--even if every node is vulnerable--robust connectivity can be maintained.

We model disjoint vulnerability by assigning every node in the network exactly one color, representing exactly that vulnerability.
The color may represent ownership, geographical location, reliance on a critical material or some other vulnerability. 
Similar to polychromatic percolation  \cite{zallen-prb1977,wierman-banach1989}, we consider the components formed by nodes of different colors separately.
We then develop a ``color-avoiding percolation'' theory which allows us to determine  the connectivity of the network when  each color (ie, the set of all nodes of a given color) is removed. 
The set of nodes that are mutually connectible under the removal of \textit{any} color comprise the color-avoiding giant component.
The existence of this component indicates whether or not the disjoint vulnerabilities can be avoided or not.



\section{Color avoiding percolation}

\begin{figure*}[htb]
\begin{center}
  \begin{minipage}{0.3\textwidth}\centering
    (a)\\
    \includegraphics[width=0.9\textwidth]{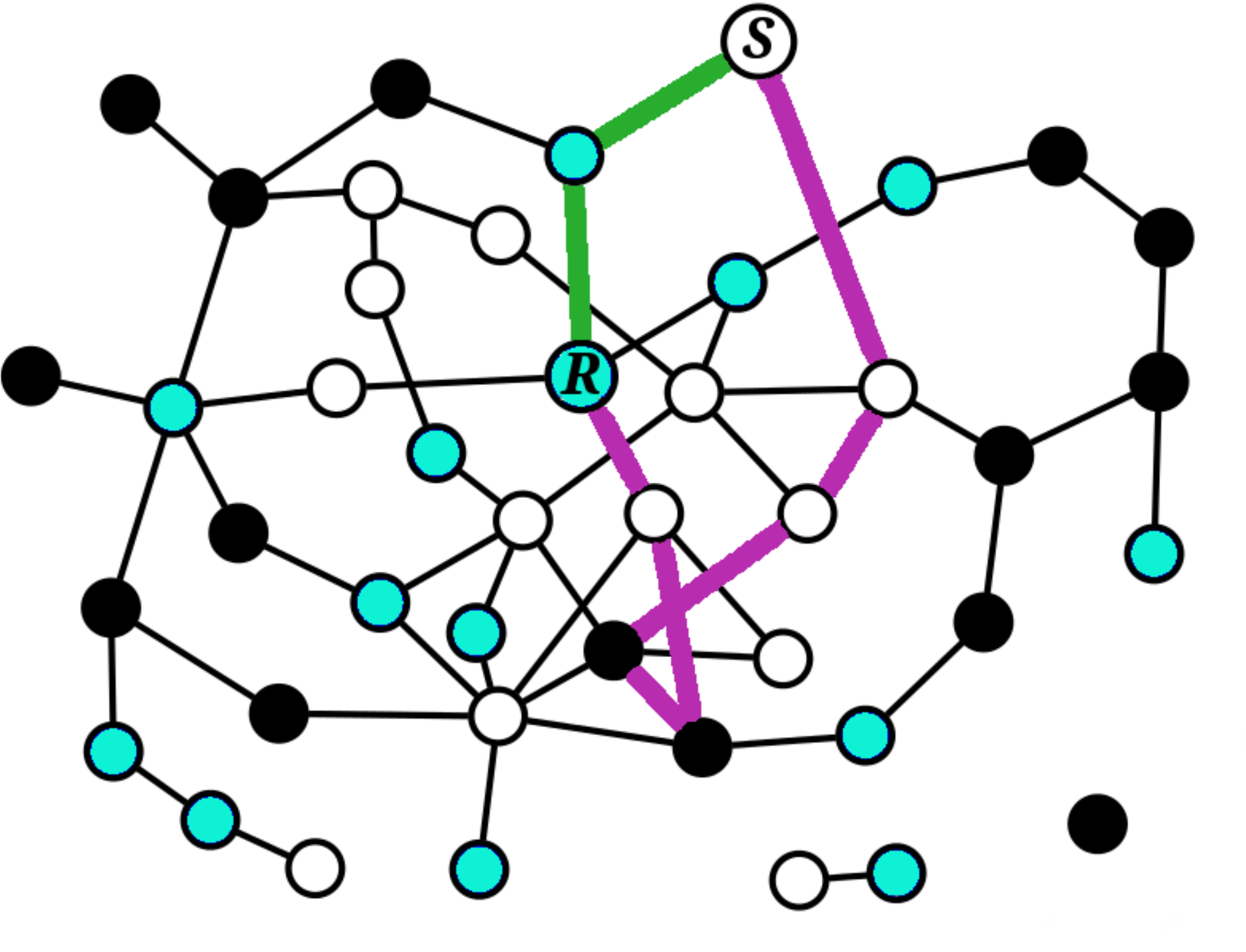}
  \end{minipage}
    \begin{minipage}{0.3\textwidth}\centering
    (b1)\\
    \includegraphics[width=0.9\textwidth]{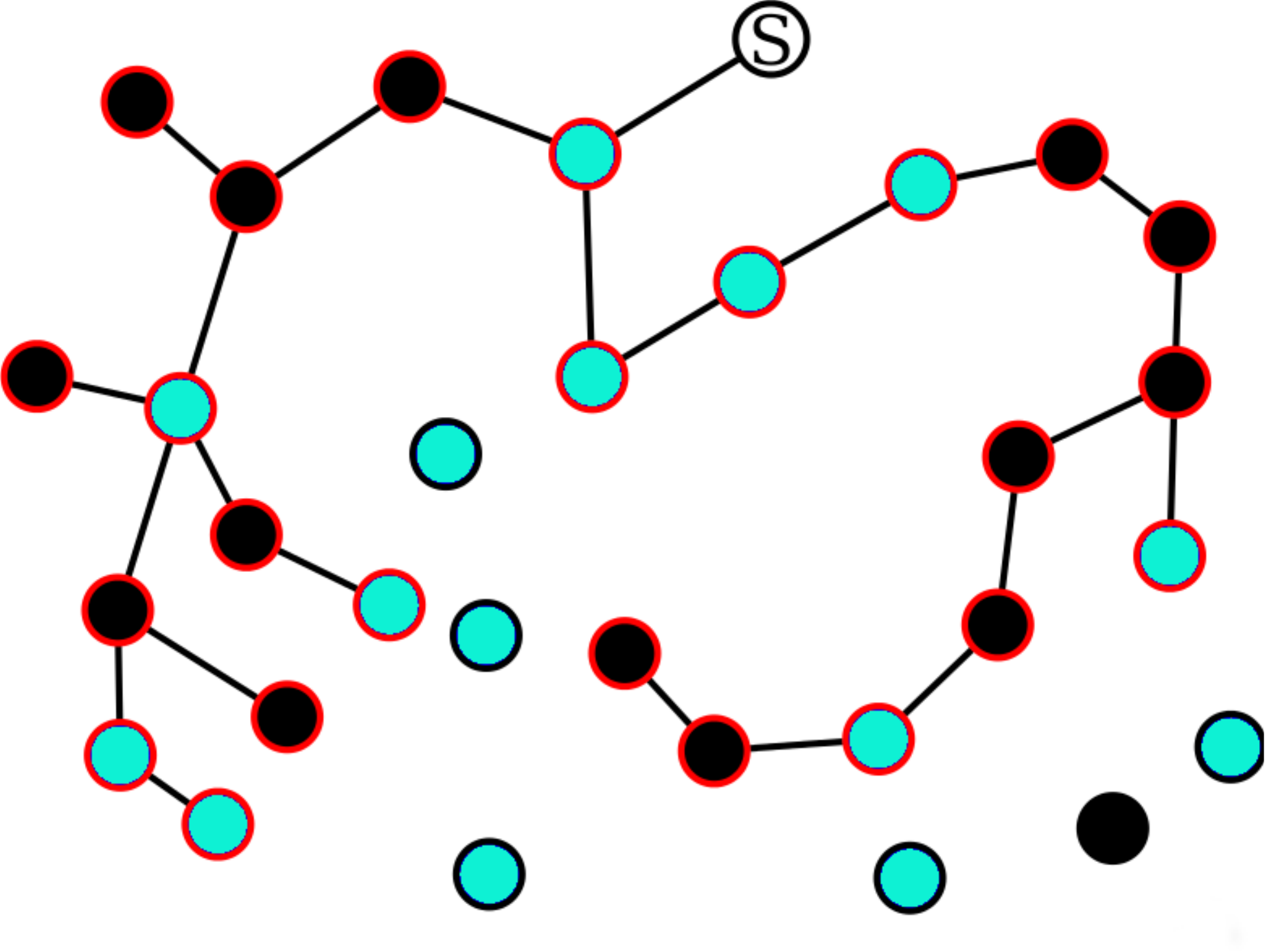}
  \end{minipage}
  \begin{minipage}{0.3\textwidth}\centering
    (b2)\\
    \includegraphics[width=0.9\textwidth]{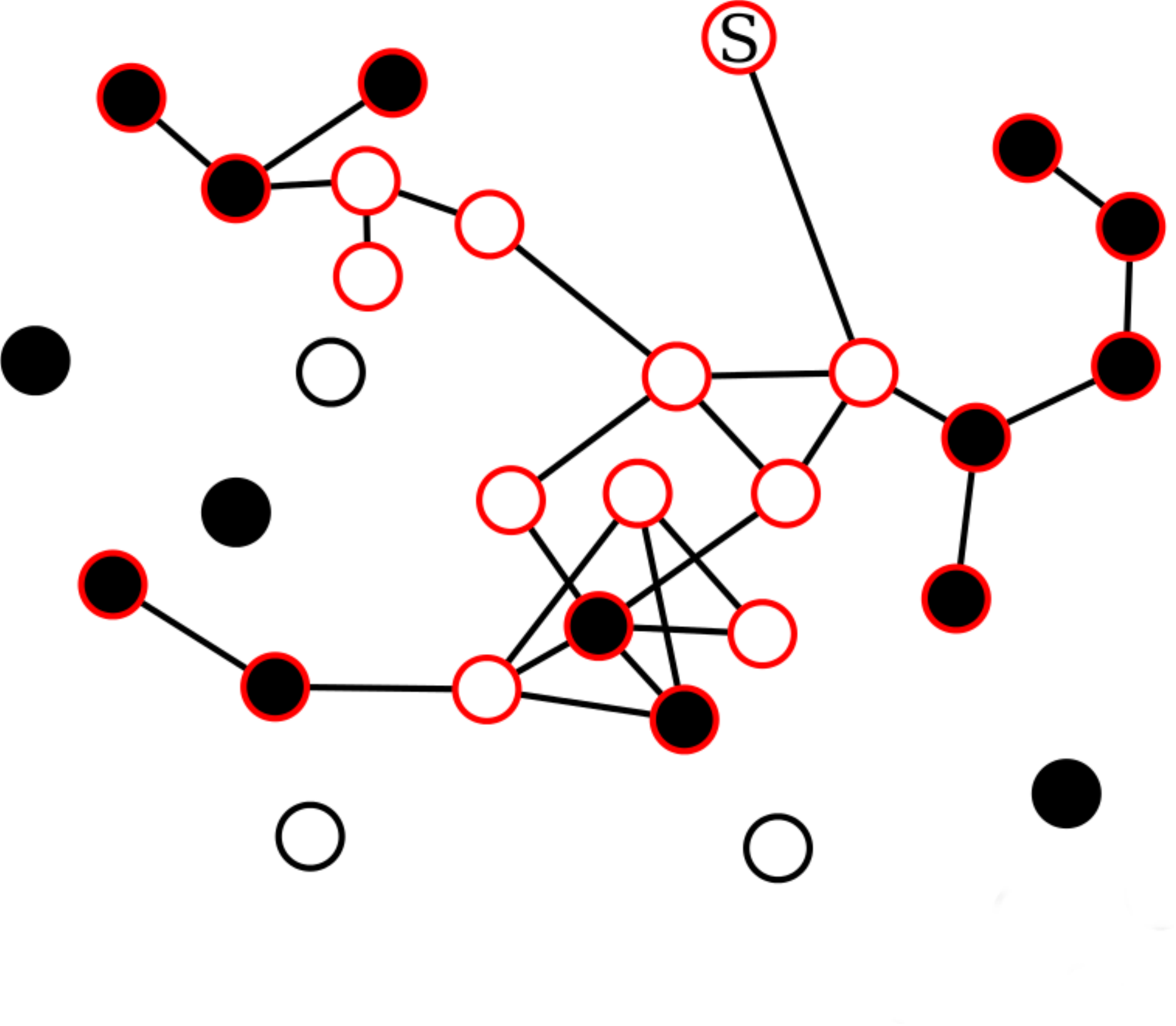}
  \end{minipage}\\
  \begin{minipage}{0.3\textwidth}\centering
    (c)\\
    \includegraphics[width=0.9\textwidth]{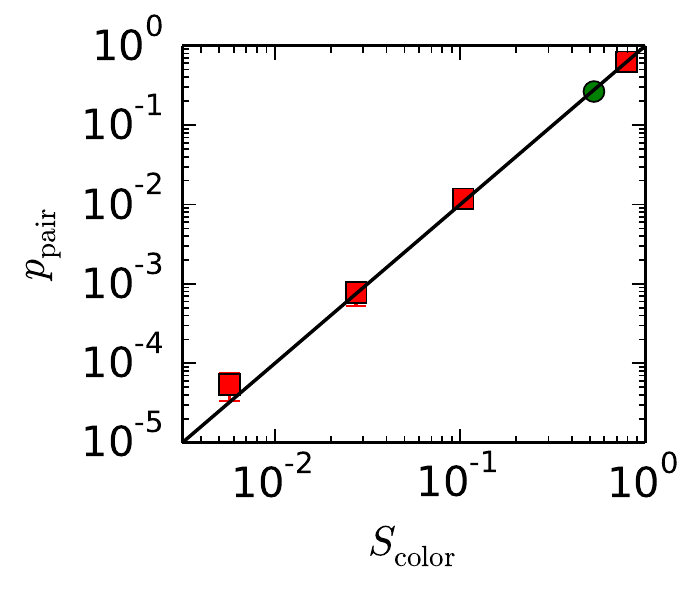}
  \end{minipage}
  \begin{minipage}{0.3\textwidth}\centering
    (b3)\\
    \includegraphics[width=0.9\textwidth]{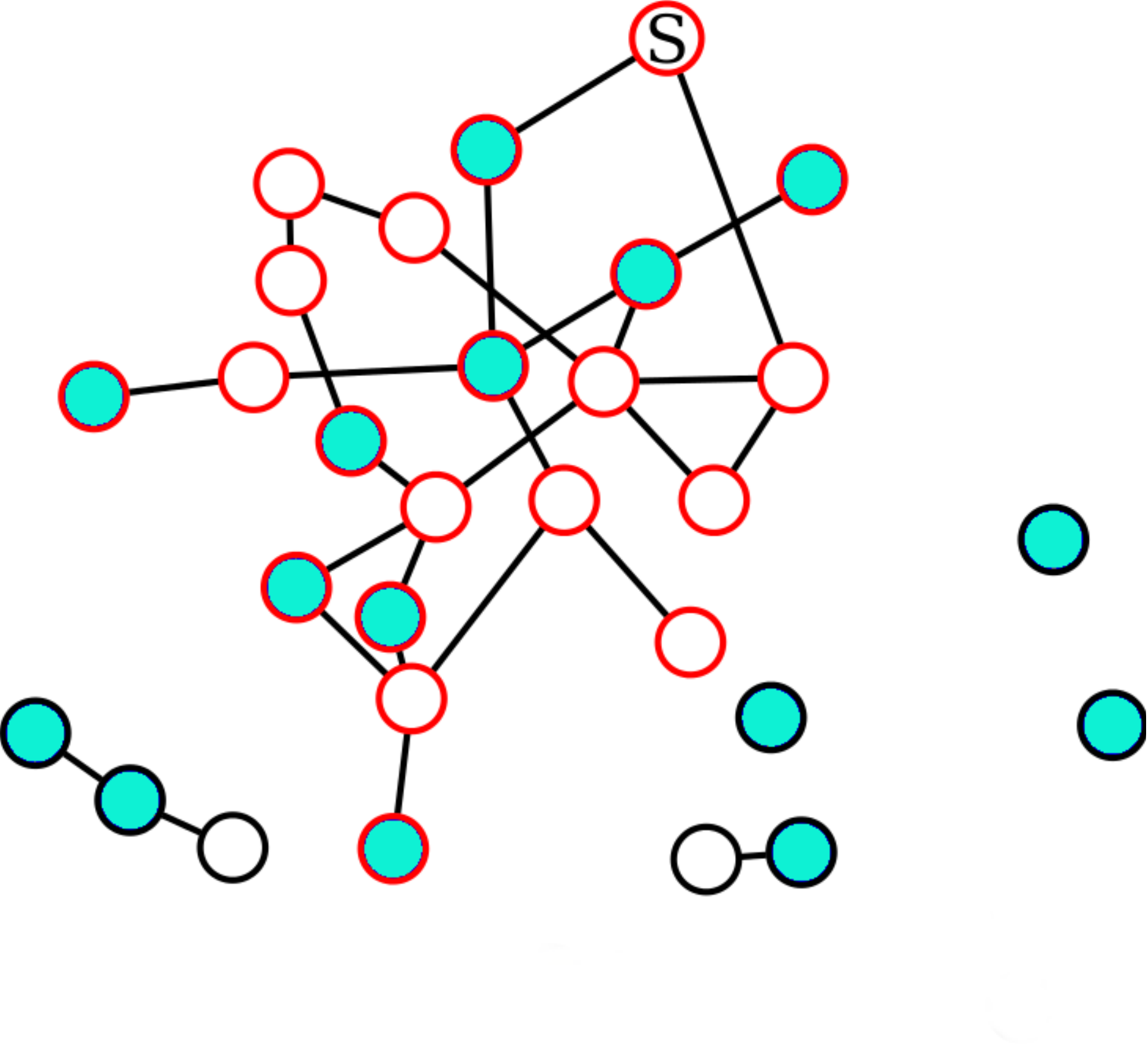}
  \end{minipage}
  \begin{minipage}{0.3\textwidth}\centering
    (b4)\\
    \includegraphics[width=0.9\textwidth]{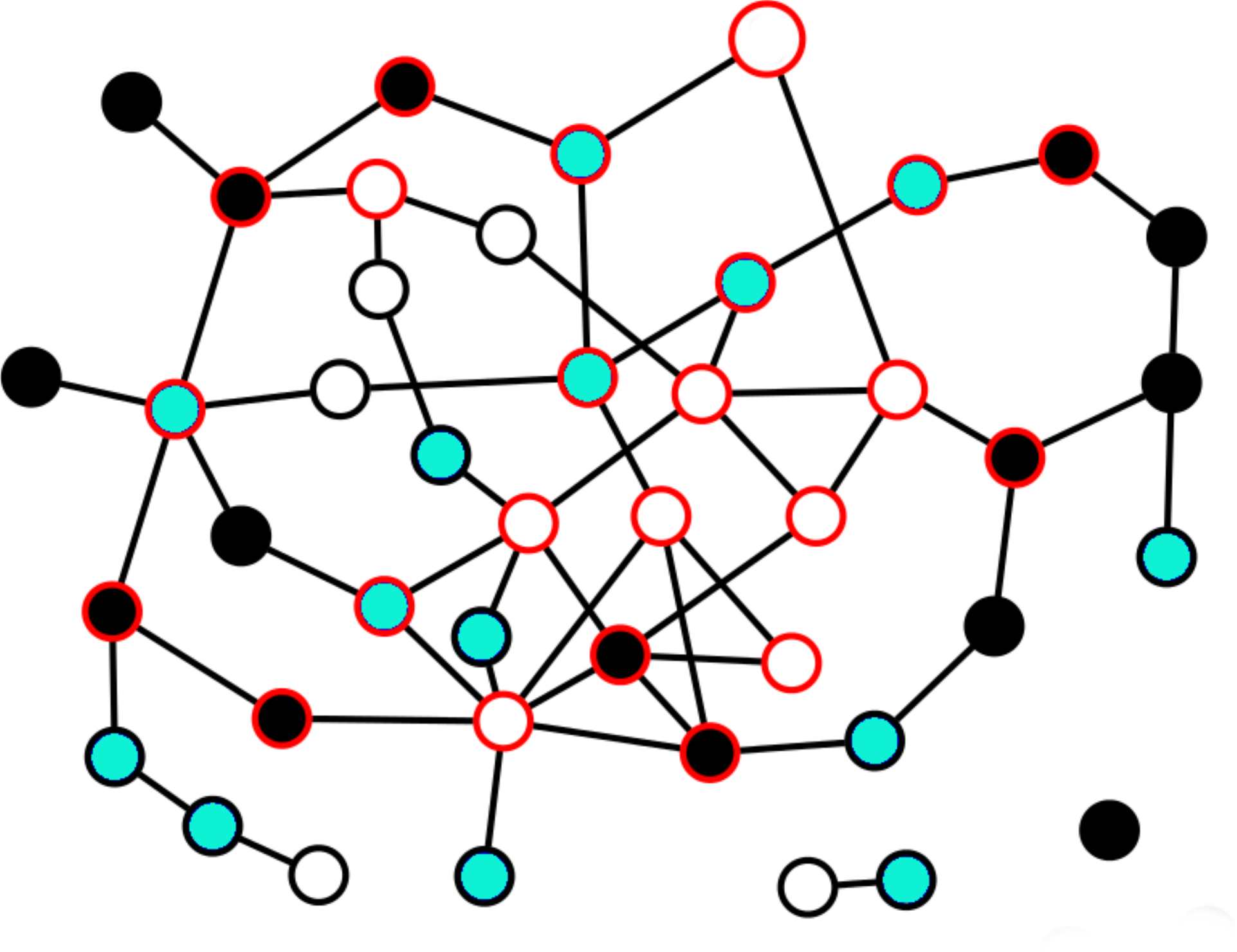}
  \end{minipage}\\
    \caption{\textbf{Illustration of color-avoiding connectivity.} 
    \textbf{(a)} In this network the sender S and the receiver R are 
    color-avoiding connected (CAC), as the green path avoids black and 
    white nodes, and the purple path avoids blue nodes. 
   \textbf{(b)} \textbf{Finding $\lcol$, the largest color-avoiding connected component.}
     \textbf{(b1)} The largest components without white ($\mathcal{L}_{\bar 1}$), 
     \textbf{(b2)} without blue ($\mathcal{L}_{\bar 2}$) and 
     \textbf{(b3)} without black nodes ($\mathcal{L}_{\bar 3}$)
    are highlighted in red, in each frame.
     \textbf{(b4)} Considering all of the nodes which are either in the largest components without each color or connected to them, we arrive at the largest CAC component, $\lcol$, the red nodes. Every pair of nodes in $\lcol$ is CAC. 
     Note that some nodes are not color-avoiding connected but are necessary to form the color avoiding components.
    \textbf{(c)} Estimation of the fraction of color-avoiding connected pairs $p_{\rm pair}$ for quenched graphs with different values $\scol$. 
    Red squares show Poisson graphs with $N=10^5$ nodes, average degrees 
${\bar k}=1.6;\,1.7;1.9;4.0$  and $C=3$ colors, the green circle 
    shows the AS network with colors representing the countries which the AS are assigned to  \cite{peixoto2014hierarchical, CaidaData}. 
    The black line indicates the case where $\scol$ accounts for all of the color-avoiding connected nodes.
    Deviations are only visible for the smallest value shown, with $\scol=570$. 
    $p_{\rm pair}$ was approximated with samples of up to $5\times 10^5$ pairs, error-bars are smaller than the symbols where not visible. 
   }
    \label{fig:avoidable_colors}
\end{center}
\end{figure*}

On a non-colored network, if node or link failures occur with a given probability, 
percolation theory can be used to determine overall connectivity \cite{cohen-book2010,newman-book2010}. 
Percolation on complex networks has a rich history \cite{boccaletti-physicsreports2006,caldarelli-sfbook2007,cohen-book2010,newman-book2010,achlioptas-science2009}.
It has been used to study the resilience of the internet \cite{cohen-2000resilience,cohen-prl2001},  its susceptibility to virus spreading \cite{pastorsatorras-prl2001} and even in probabilistic routing algorithms~\cite{sasson2003probabilistic}.
It has also been used to understand word-of-mouth processes in social networks~\cite{goldenberg-physa2000,solomon-physa2000}, and the robustness of many biological networks including neural networks \cite{breskin-prl2006}, metabolic networks \cite{smart-pnas2008} and mitochondrial networks \cite{aon-pnas2004}.
Here we develop a new framework based on percolation theory but not reducible to any previous percolation problems.
In this framework, connectivity corresponds to the ability to avoid disjoint vulnerabilities via multiple paths.

We begin with an undirected unweighted network $G$ with $N$ nodes and adjacency matrix $A_{ij}$. 
Every vertex $i$ is assigned a color $c_i\in\{1,2,\dots,C\}$, where $C$ denotes the total number of colors. 
Faced with the possible vulnerability or insecurity of all nodes of a single color, we seek 
a set of paths between two nodes such that \textit{no color} is required for \textit{all paths}.
In non-colored graphs, a single path provides connectivity and in $k$-core percolation any $k$ paths are sufficient \cite{dorogovtsev-prl2006,goltsev-pre2006}.
We now define a pair of nodes as `` color-avoiding connected''  (CAC) if, for every  color $c$, there exists a path connecting this pair and \textit{avoiding} all nodes of color $c$.
We assume that the source and target themselves are secure, and their colors are not included in the calculation of color-avoiding connectivity. 
The paths are not necessarily unique: often one path can avoid multiple colors (see Figure~\ref{fig:avoidable_colors}a).
However, if $C$ paths cannot avoid all $C$ colors, then the source and target require one of the colors to be connected and adding more paths will not help.
Since avoiding disjoint vulnerabilities through multiple paths is a feasible strategy only if a giant CAC component exists,
we do not address optimal path problems but rather focus on the properties of CAC components.




Formally, we define a ``color-avoiding connected component'' as a maximal set of nodes, 
where every node pair in the set is color-avoiding connected. 
Several examples of CAC components are shown in Fig. \ref{fig:avoidable_colors}b and Supp. Fig 1.
Note that there are nodes which are not themselves part of the CAC component 
but are necessary for the color-avoiding connectivity of nodes which are in the component.
This occurs, for example, when all of the neighbors which lead from a node to the CAC component are of the same color.
In such a case, the node itself is not CAC to the system as a whole because it must pass through nodes of a certain color before it can reach elsewhere.  
However, in general, this node will still be necessary to form paths which avoid other colors.
The fact that non-CAC nodes may be needed to create overall system color-avoiding connectivity is 
one indication that a new kind of percolation theory is needed to uncover this hidden structure.

By studying the largest CAC component, we obtain a clear quantitative measure of the feasibility of  multiple paths to avoid disjoint vulnerabilities and information on where those paths should be routed.
Furthermore, this gives us a way to measure the effect of changes in network topology, link density and color distribution.

To find the largest set of color-avoiding connected nodes in any network with any color distribution, we propose the following algorithm.
First, for every color $c$, we delete all nodes with color $c$ 
and find the largest component in the remaining graph, $\mathcal{L}_{\bar c}$. 
Next, we define $\mathcal{L}_{\rm color}$ as the set of nodes which, for every color $c$,  are either (a) in $\mathcal{L}_{\bar c}$ or (b) have at least one link to it.
Condition (b) represents the assumption that the color of the source and target are not included in the calculation.
If we only used condition (a), the calculation of $\mathcal{L}_{\rm color}$ from $\{\mathcal{L}_{\bar c}\}$ would be equivalent to the calculation of the mutual giant component in interdependent \cite{buldyrev-nature2010} or multiplex networks \cite{baxter-prl2012,boccaletti-physicsreports2014} and the result would always be an empty set because every node has some color $c'$ and  is therefore not a member of $\mathcal{L}_{\bar c'}$.
In Figure~\ref{fig:avoidable_colors}b, we illustrate this method and further technical details are discussed in Supp. Sec. 1.A.


It is possible that $\mathcal{L}_{\rm color}$ does not represent the overall color-avoiding connectivity of the system due to smaller components.
However, if $\lcol$ scales with system size and the smaller color-avoiding connected components do not, 
then in the limit of large systems the overall color-avoiding connectivity is determined by $\lcol$ just like the overall connectivity is determined by the size of the giant component in non-colored graphs.
With $\scol$ defined as the fraction of the total nodes which are in  $\mathcal{L}_{\rm color}$ 
and $p_{\rm pair}$ defined as the total fraction of color-avoiding connected pairs among all node pairs,
we can test if $\lcol$ accounts for the bulk of color-avoiding connectivity.
In Figure~\ref{fig:avoidable_colors}c we see that color-avoiding connectivity is indeed dominated by $\mathcal{L}_{\rm color}$ for random and real-world networks. 
When $\scol$ is small, non-giant clusters and the trivial color-avoiding connectivity 
which accompanies individual links leads to deviations between $p_{\rm pair}$ and $\scol$ but these deviations rapidly disappear as the sytem size increases. 
 This validates the treatment of $\mathcal{L}_{\rm color}$ as a proxy for color-avoiding connectivity.
 We proceed  to develop analytical results based on percolation theory for random networks.
 

\section{Analytic theory for random networks}
\begin{figure*}[htb]
\begin{center}
\begin{minipage}{0.32\textwidth}
(a)\\
\includegraphics[width=0.95\textwidth]{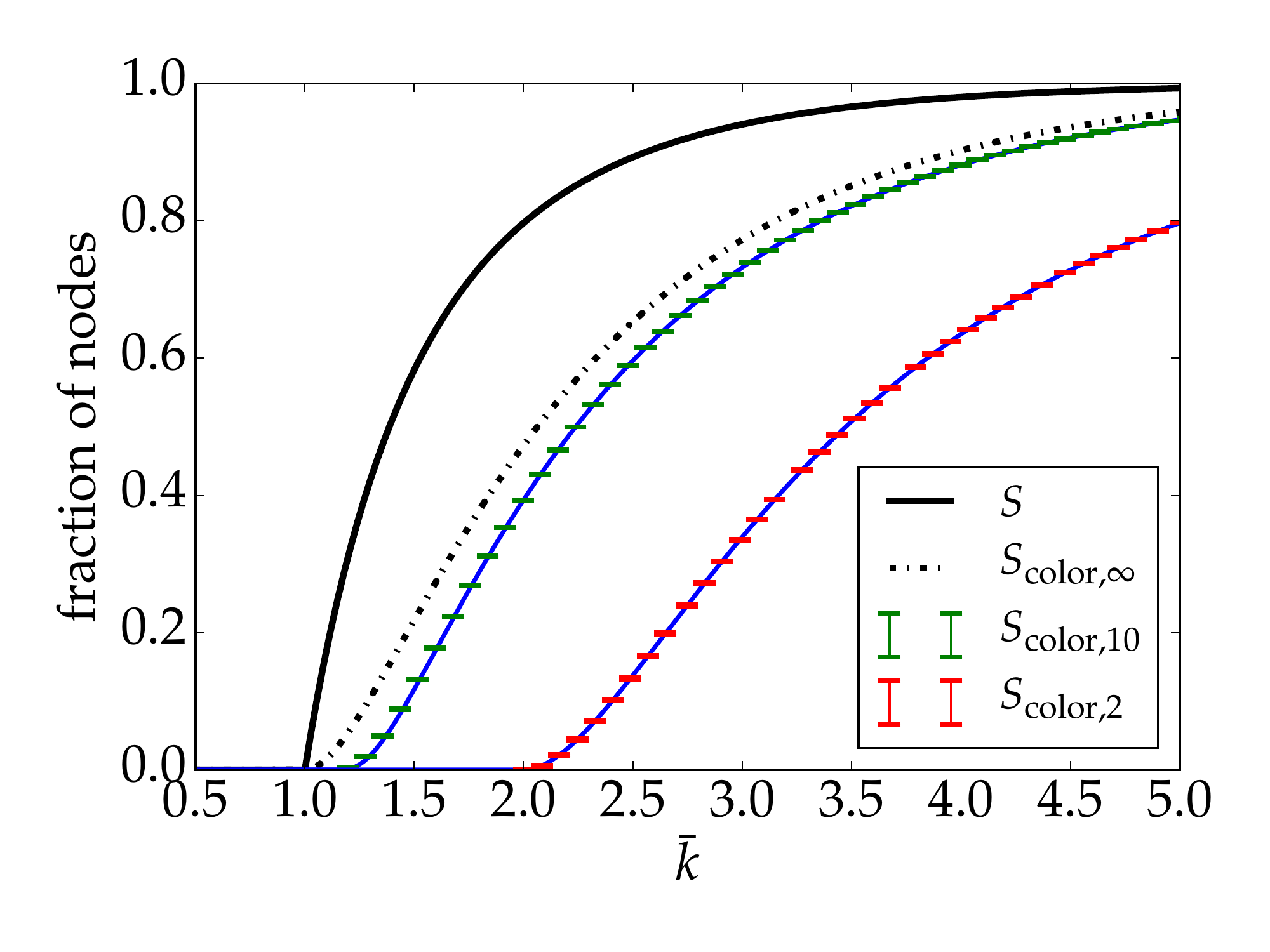}
\end{minipage}
\begin{minipage}{0.32\textwidth}
(b)\\
    \includegraphics[width=0.95\textwidth]{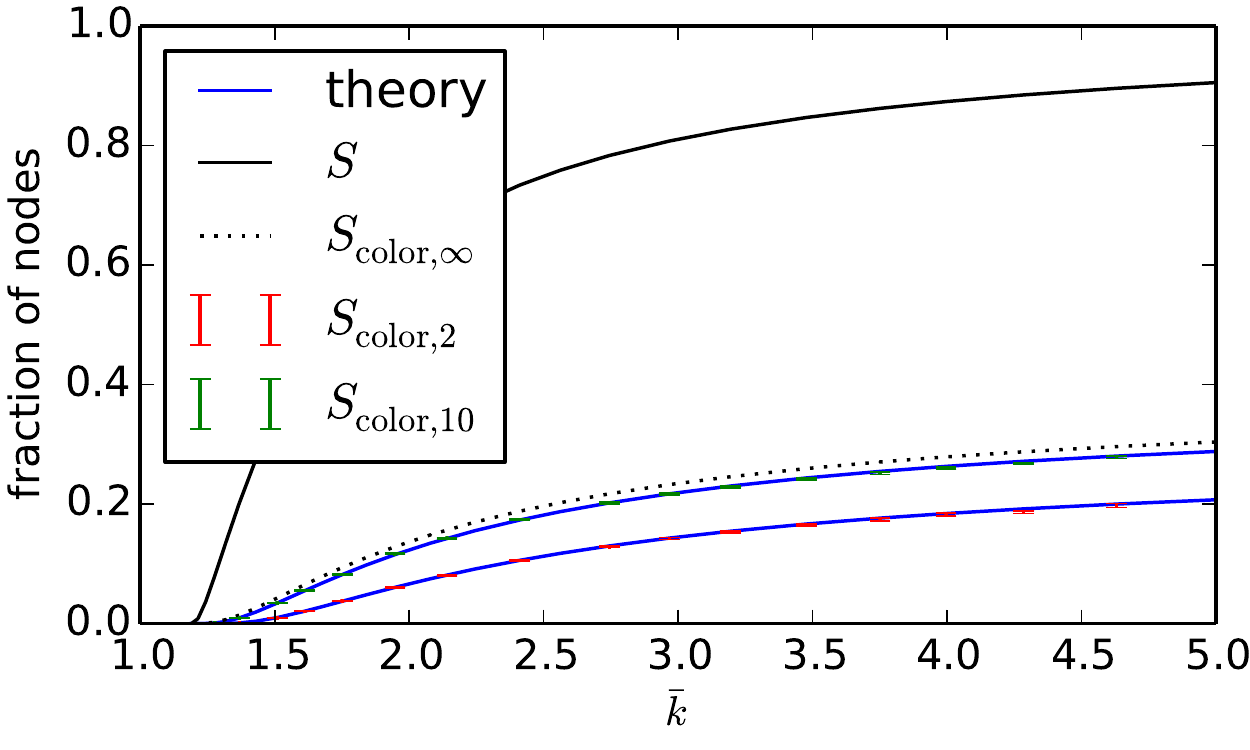}
\end{minipage}
\begin{minipage}{0.32\textwidth}
(c)\\
    \includegraphics[width=0.95\textwidth]{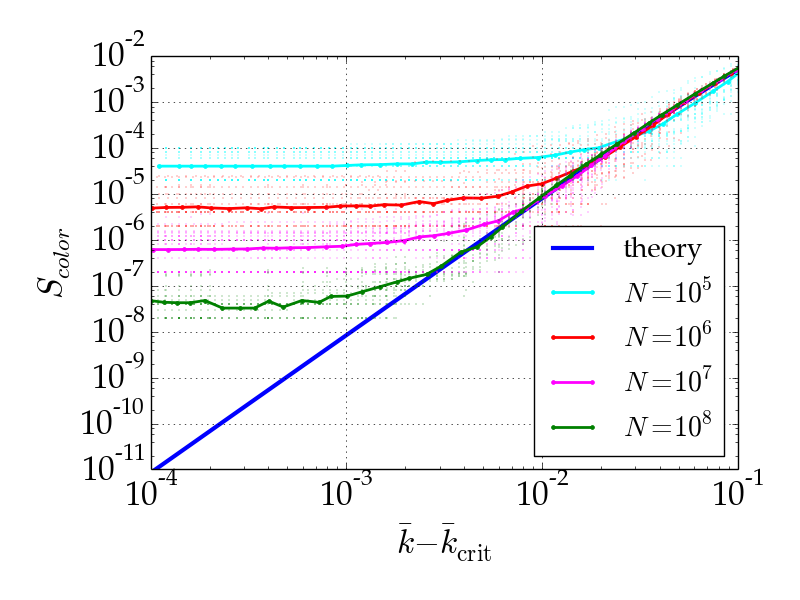}
\end{minipage}
    \caption{\textbf{Size of the giant color avoiding component $\scol$ in random networks with uniformly distributed colors.}
Dependence of $S_{\rm color}$ on average degree $\kk$ \textbf{(a)} for \er  networks and \textbf{(b)} scale-free networks with different 
    numbers of colors.
    Error bars are shown but barely visible for networks of size $N=10^6$.
    The blue lines show the corresponding analytical results. 
    For comparison, we include the giant component size of standard percolation $S$ (black solid) and 
    the limiting case of a system with an infinite number of colors, $S_{{\rm color},\infty}$ (black dashed).
    As mentioned in the text, $S_{{\rm color},\infty}$ is the same as the giant component in 2-core percolation.
    \textbf{(c)}: Critical exponent and finite size scaling for \er  networks with $C=3$.
    Note that in the critical region the theory and simulations show a slope of almost exactly 3 as predicted by Eq. \ref{eq:critical_params}.
    Finite size scaling is shown with the results of $>150$ realizations per size plotted individually and averaged.
    }
    \label{fig:analytics}
\end{center}
\end{figure*}

For the analytical treatment we use the annealed approximation of networks of size $N$ described through the configuration model \cite{newman-book2010}, in which a degree distribution $p(k)$ is a conserved quantity from which an ensemble of network realizations is drawn. 
For a more comprehensive treatment see the supplementary information.

Every node $i$ is assigned a color $c_i\in \{1,2,\dots,C\}$. 
The analytic framework presented here assumes that the colors are distributed uniformly at random.
Hence, the color sequence $\{c_i\}$ has probability $\prod_i r_{c_i}$ with the color frequencies $r_c$.

We calculate $S_{\rm color}$ in the limit of $N\to \infty$ as the probability that a
single node belongs to $\mathcal{L}_{\rm color}$.  
Because $\mathcal{L}_{\rm color}$ is a subset of the regular giant component by construction,
we begin by obtaining the solution for standard percolation on random graphs \cite{erd-1959random,newman-2001random,newman-book2010}.
The size of the giant component in a non-colored random graph is $S = 1 - g_0(u)$ where $g_0(z)=\sum p_k z^k$ is the generating function of the probability distribution $p_k$. 
$u$ is the probability that a node is not connected to the 
giant component over one particular link and is computed as the solution of $u=g_1(u)$, where $g_1(z)=g_0'(z)/g_0'(1)$
is the generating function of excess degree \cite{newman-book2010}. 
Second,
we let $\kappa_c$ be the expected number of a randomly chosen node's neighbors of color $c$ which are connected to the giant component of standard percolation.
Considering $\kappa_c$ for all colors, we obtain the vector ${\vec \kappa}=(\kappa_1,\dots,\kappa_C)$ with $k'=\sum_c \kappa_c$ being the total number of links to the normal giant component.
Third, the  conditional probability $P_{\vec \kappa}$ that the links suffice to connect to $\mathcal{L}_{\rm color}$, given that they belong to distribution $\vec{\kappa}$ and that they already belong to the normal giant component, is:
\begin{align} \label{eq:pveckappa}
P_{\vec \kappa} &= \prod_{c=1}^{C}\left(1 - U_{\bar c}^{k' - \kappa_c}\right),\\
U_{\bar c} &= 1 - \frac{1-u_{\bar c}}{(1-u)(1-r_c)},\label{eq:U_c}
\end{align}
in which $U_{\bar c}$ denotes the conditional probability that a link 
fails to connect to $\mathcal{L}_{\bar c}$ given that it does connect to the normal giant component via a node having a color $c'\neq c$. We define $U_{\bar c}=1$ if $u=1$. The probability $u_{\bar c}$ that a single link does not connect to a giant $\mathcal{L}_{\bar c}$ 
is calculated with $u_{\bar c} = r_c + (1-r_c) g_1(u_{\bar c})$ (site percolation with a surviving fraction of nodes of $1-r_c$ \cite{newman-book2010}). 
%
Combining these terms, we obtain a formula for $\scol$:
\begin{equation}
S_{\rm color} = \sum_{k=0}^{\infty}p_k \sum_{k'=0}^{k} B_{k,k'} 
\sum_{\kappa_1,\dots, \kappa_C=0}^{k'} M_{k',\vec \kappa} 
P_{\vec \kappa},\label{eq:s_color} 
\end{equation}
where the binomial factor $B_{k,k'}$ (Supp. Eq. S7) accounts for the probability that out of 
$k$ links $k'$ links connect to the normal giant component. The multinomial factor 
$M_{k',\vec \kappa}$ (Supp. Eq. S8) gives the multinomial probability of having the color 
distribution ${\vec \kappa}$ among the neighbors belonging to the normal giant component. 



To obtain a closed-form solution for $\scol$, we now assume that every color occurs with equal probability: $r_c = 1/C$. 
With $U_{\bar 1}=U_{\bar c}$ being identical for 
all colors 
we have  (Supp.  Eq. S20):
\begin{align}
S_{{\rm color},C} &=  \sum_{j=0}^C (-1)^j {C \choose j} \times\nonumber \\
&\, \times g_0\left\{u+(1-u)\left[\frac{j}{C}U_{\bar 1}^{j-1} + \frac{C-j}{C}U_{\bar 1}^{j}\right]\right\}.
\end{align}

We now discuss the limiting cases $C=2$ and $C\to \infty$. 
The result for two colors can be simplified to (Supp. Eq. S17)
\begin{align}
S_{{\rm color},2} &=1-2 g_0(u_{\bar 1})+g_0(2 u_{\bar 1}-1)
\end{align}
which directly depends on $u_{\bar 1}$ only.
As the number of colors tends to infinity, standard percolation \textit{is not} recovered and $\scol$ remains smaller than the relative size of the giant component $S$ and in fact $S_{{\rm color},\infty}$ is identical to the giant component in $k$-core percolation with $k=2$ \cite{dorogovtsev-prl2006,goltsev-pre2006}.
The reason that $S_{{\rm color},\infty}$ is equivalent to 2-core percolation is that--even if every node is a different color--if a node were connected via only one link,  it would not be able to avoid the color of its sole neighbor.
We demonstrate this directly by deriving an asymptotic form for $\scol$ as $C\rightarrow\infty$ (Supp. Eq. S23):
\begin{equation}\label{eq:scolinf}
S_{{\rm color},\infty} = S - \left. (1-u)\frac{dg_0(z)}{dz}\right| _{z=u}
\end{equation}
which is the same result as in 2-core percolation.
In Fig. \ref{fig:analytics}a we see that $S_{{\rm color},C}$ comes close to $S_{{\rm color},\infty}$ even for $C=10$, indicating that even moderate color diversity comes close to the infinite color case. 

We now discuss graphs with broad degree distributions with $p_k\sim k^{-\alpha}$ ($k>0$)
and generating functions $g_0(z)={\rm Li}_{\alpha}(z)/\zeta(\alpha)$ and 
$g_1(z)={\rm Li}_{\alpha-1}(z)/[z\zeta(\alpha-1)]$, with ${\rm Li}_{\alpha}(z)$ the 
polylogarithm function. 
In Figure~\ref{fig:analytics}b we see results for $C=2$ and $C=10$ depending on 
the average degree $\kk = \zeta(\alpha-1)/\zeta(\alpha)$ \cite{newman-book2010}. The limiting cases are 
diverging $\kk$ for $\alpha=2$ and $\kk=1$ for $\alpha\to\infty$. 
We see that $\kk_{\rm crit}$ is not strongly affected by the number of colors but that the size of the giant CAC component is substantially smaller than in the case of \er  networks (see Figure \ref{fig:analytics}a-b).
The critical connectivity can be calculated using 
Cohen's criterion for site percolation~\cite{cohen-2000resilience}. 
With  the fraction $1-r_c$ of nodes surviving random removal, we obtain
$1-r_c=1-1/C=\kk/(\left<k^2\right>-\kk)$.
Since $\left<k^2\right> = \zeta(\alpha-2)/\zeta(\alpha)$, we have 
$\zeta(\alpha-2)/\zeta(\alpha-1)=1+C/(C-1)$. 
Accordingly $\kk\approx1.254$ for two colors, and it converges to $\kk\approx1.195$ for $C\to \infty$. 

We find that \er~networks are more color-avoiding connected than scale-free networks of equal average degree, the opposite of the results for resilience to random failures \cite{albert2000error,cohen-prl2001,newman-book2010}. 
This follows from the difference in the 2-core envelopes; compare Figs. \ref{fig:analytics}a and \ref{fig:analytics}b.

\begin{figure*}
\centering
  \begin{minipage}{0.37\textwidth}\centering
    \includegraphics[width=\textwidth,trim=30 15 20 75,clip=true]{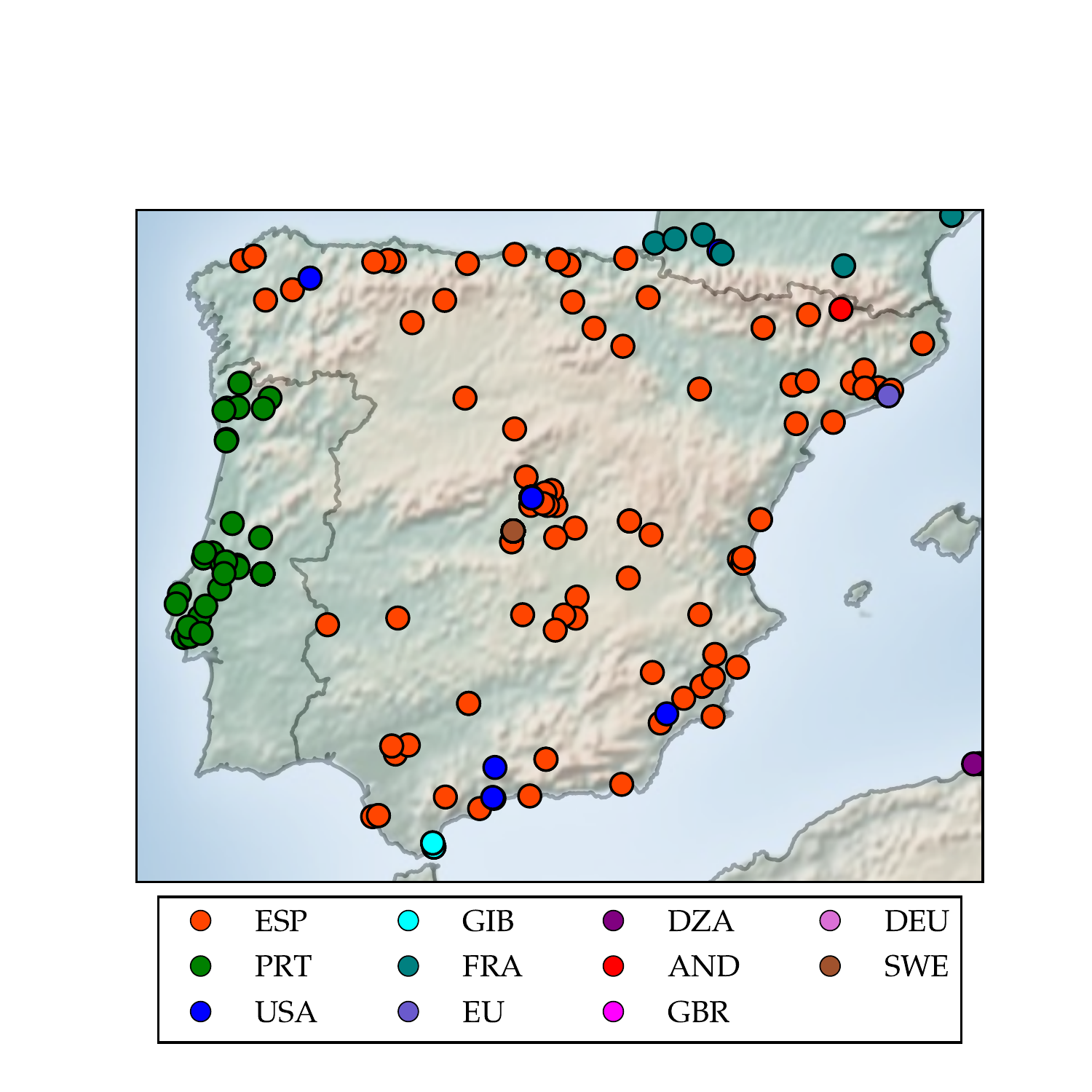}
    (a)\\
  \end{minipage}
  \begin{minipage}{0.37\textwidth}\centering
    \includegraphics[width=\textwidth,trim=30 15 20 75,clip=true]{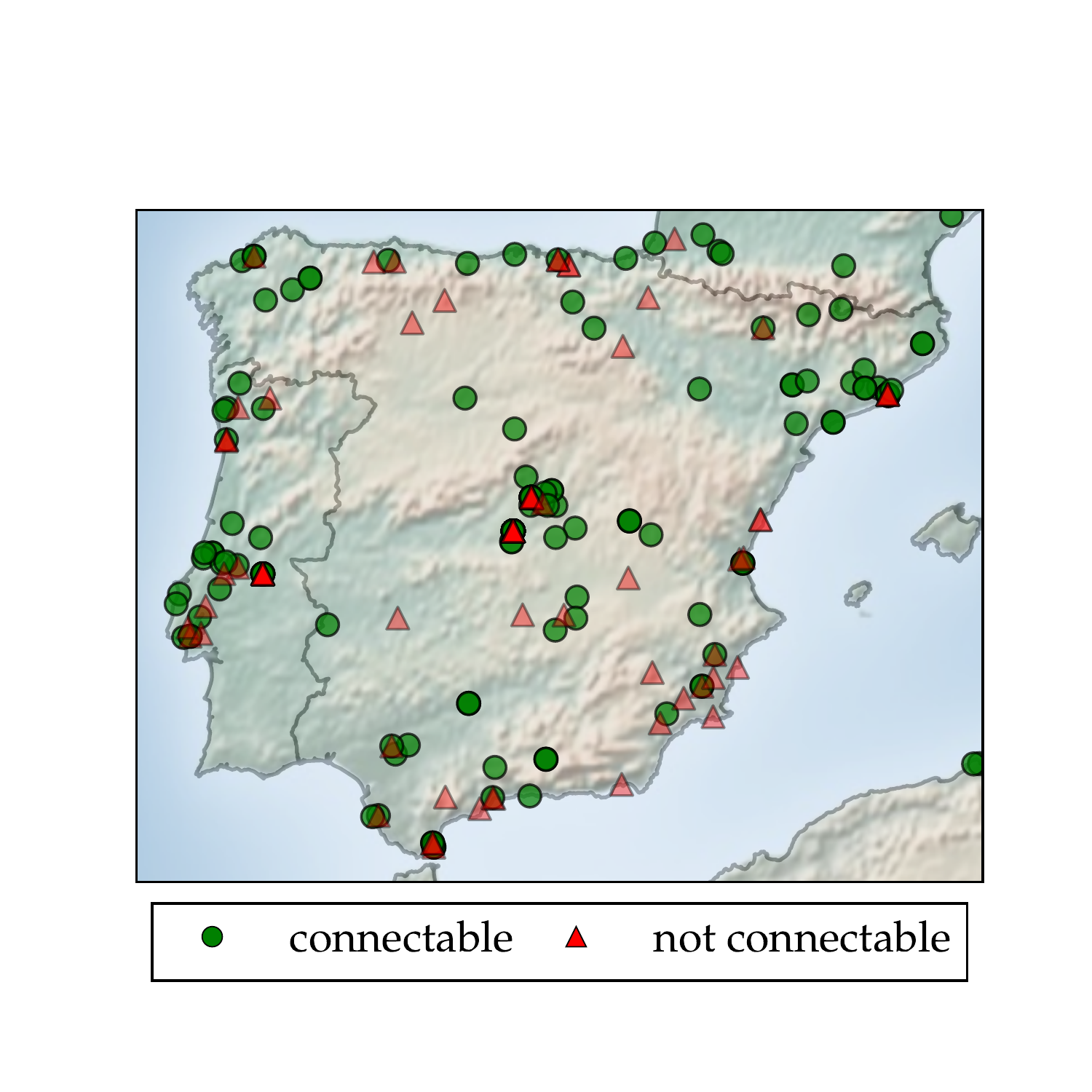}
    (b)\\
  \end{minipage}
  \begin{minipage}{0.23\textwidth}\centering
    \includegraphics[width=\textwidth]{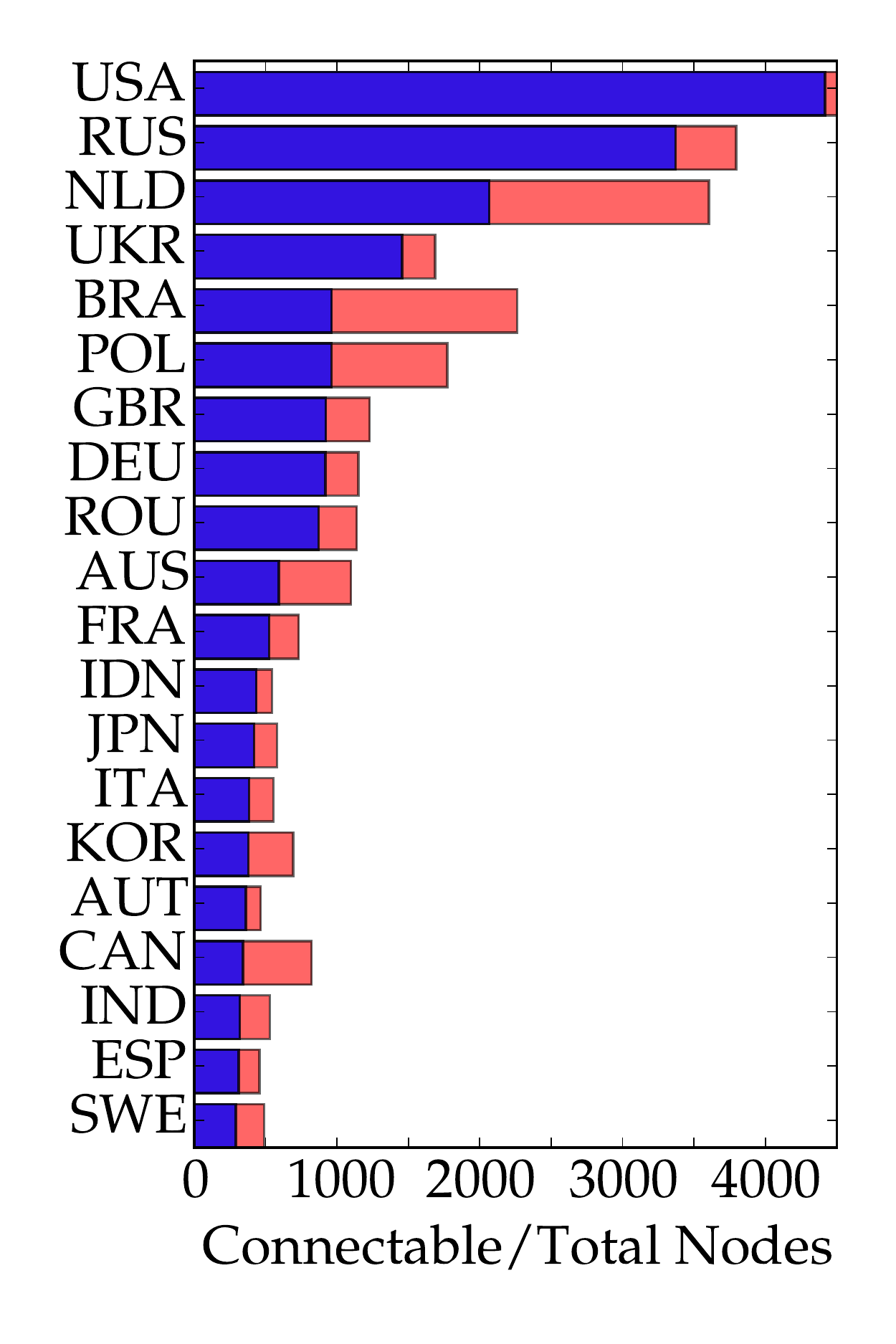}
    (c)\\
  \end{minipage}
\caption{\textbf{Color-avoiding connectivity of the AS-level internet.} 
\textbf{(a)} Here we show the routers of the AS-level internet in the Iberian peninsula as a disjointly vulnerable network with the colors determined by the country to which the router is registered. 
\textbf{(b)} Using these colors, we calculate the largest color-avoiding connected component.  The green nodes are members of this set while the red are not.  This means that these routers can take advantage of multiple paths to maintain security, as desribed in the main text.
\textbf{(c)} This shows the number of CAC routers (nodes in $L_{\rm color}$) compared to total number of routers for the top 20 countries worldwide, in terms of total number of AS routers registered to that country. Data for the US has been trunctated for visibility, the total number of AS routers is 17690.
We use a symmetrized version of the network of~\cite{peixoto2014hierarchical} which was generated using data from the CAIDA project~\cite{CaidaData} up to December 2013. }
\label{fig:AS_data}
\end{figure*}
\section{Critical phenomena}

We now turn to the critical behavior of $\scol$ in \er-graphs with $C$ uniformly distributed colors. 
Similar to standard percolation, we find that the size of the largest color-avoiding connected component $\scol$ undergoes a phase transition at a specific $\kk_{\rm crit}$, which is now determined by the number of colors 
see Figure~\ref{fig:analytics}. 
For $\kk < \kk_{\rm crit}$, color-avoiding connectivity is confined to clusters of finite size (zero in the limit of large $N$) and for $\kk > \kk_{\rm crit}$ there is a largest color-avoiding connected component $\scol$ which scales with system size.
We find that the value of $\kk_{\rm crit}$ decreases as $C$ increases and approaches the standard percolation threshold as $C\rightarrow\infty$. 
Since color-avoiding connectivity requires that the giant component not be destroyed after the removal of any single color, we require that $\kk_{crit}^{ER} = \kk_{crit} \frac{C-1}{C}$ where $\kk_{crit}^{ER}=1$ is the percolation threshold for ER graphs and $\frac{C-1}{C}$ is the fraction of links remaining after the removal of $1/C$ nodes.
Therefore $\kk_{\rm crit} = C/(C-1)$. 

To discuss the scaling and critical exponents, we return to the definition of $P_{\vec\kappa}$, Eq. \ref{eq:pveckappa}. 
We consider the region close but above $\kk_{\rm crit}$ by defining $\varepsilon\equiv 1-U_{\bar 1}\approx C (\kk-\kk_{\rm crit})$ which holds as long as $(\kk-\kk_{\rm crit})\ll 1/C$ (Supp. Eq. S27). 

We analyze the behavior of $P_{\vec \kappa}$ for small $\varepsilon$ by expanding 
$(1-(U_{\bar 1})^{k'-\kappa_c})\approx (k'-\kappa_c)\varepsilon$. 
Plugging this approximation in to Eqs. \ref{eq:pveckappa} and \ref{eq:s_color} we obtain:
\begin{align}
\label{eq:scaling_relation}S_{\rm color} &\propto ({\bar k}-{\bar k}_{\rm crit})^{\beta}\\
\label{eq:critical_params}\beta&=C,\quad {\bar k}_{\rm crit} = C/(C-1).
\end{align}
We confirm the value of $\kk_{\rm crit}$ and the scaling of $\scol$  numerically in Figure \ref{fig:analytics}c for $C=3$ colors.
As $C\rightarrow\infty$, we need to resolve the seeming contradiction of 
a divergent critical exponent $\beta=C$ and convergence towards $S_{{\rm color},\infty}$ as it appears in Eq. \ref{eq:scolinf}. 
For ER networks we show (Supp. Eq. S31) that $S_{\rm color,\infty}\propto ({\bar k}-1)^2$ 
for $\kk$ near 1, implying $\beta=2$. 
The reason that we do not observe $\beta\to \infty$ as described in Eq. \ref{eq:critical_params} is that
the approximation used to obtain Eq. \ref{eq:scaling_relation} is only valid in a critical region defined as
$(\kk-\kk_{\rm crit})\ll 1/C$. 
As $C\rightarrow\infty$, $\scol$ increases with the high exponent $\beta=C$.
However, the shrinking critical region overpowers the diverging critical exponent and $\scol\sim 0$ takes on unobservably small values and  crosses over to $\beta=2$ scaling outside the critical region.

\section{Applications}

One immediate application of our framework is to secure communication in a network with no trusted nodes.
Assuming $C$ router owners, each of whom eavesdrops on its routers traffic, we can securely communicate if messages are split with a \textit{secret sharing} protocol \cite{blakley1899safeguarding,shamir1979share,dolev-acm1993} and transmitted along multiple color-avoiding paths.
The nodes which can take advantage of this method are exactly the elements of the largest CAC component.

To study the hidden CAC structure of the internet, we use a symmetrized version of the AS-level internet prepared by \cite{peixoto2014hierarchical} which was generated using data from the CAIDA project~\cite{CaidaData} up to December 2013. 
We then color every router according to the country to which the router is registered, reflecting the assumption that every country is eavesdropping on its traffic but that no countries share information (Fig. \ref{fig:AS_data}a).
Using the algorithm for finding the largest CAC component, we can determine which nodes are color-avoiding connectable and which are not (Fig. \ref{fig:AS_data}b).

We find that overall $26228$ out of $49743$ ($\approx52.73\%$) of the routers are in the largest CAC component and that this accounts for the vast majority of CAC connected nodes (Fig. \ref{fig:avoidable_colors}c).
However, we also find that these results vary greatly from country to country.
For instance, only $25\%$ of the routers registered to the United States are in the largest CAC component compared to $89\%$ of routers registered to Russia (Fig. \ref{fig:AS_data}).
This is partially due to the density of routers in the US which is much higher than Russia and indicates that US eavesdroppers have far greater capacity to intercept communication than their Russian counterparts.

In economic trade networks, it is common that a single firm controls many others \cite{vitali-plosone2011} but each firm is controlled by only one owner.
The vulnerability to correlated failures or malicious activities can undermine the overall system robustness, if they are sufficient to disrupt the global color-avoiding connectivity.
We thus add color-avoiding connectivity to the concerns regarding systemic risk and government regulation of mergers and acquisitions \cite{battiston-sreps2012,tessone-jstatphys2013}.

In epidemiology, many diseases spread via different strains, and individuals may become immune after recovery \cite{masuda-jtheoretbio2006}.
Coloring nodes by strain,  color-avoiding percolation can be used to evaluate the population's susceptibility 
to a multi-strain infection.

\section{Discussion}

We have presented here the first systematic study of disjoint vulnerabilities in complex networks and a way to maintain network robustness by utilizing multiple paths.
We have shown that even a small diversity of colors can enable color-avoiding connectivity to a large fraction of nodes in a random network but that in real-world networks, uneven distribution of vulnerabilities can undermine this effect.
The framework and metrics uncover a hidden structure that underlies any complex network with nodes that can be partitioned by their susceptibility to an external threat and can be used to devise new network design principles and protocols for improving robustness through redundancy.

\section*{Author Contributions}
All authors contributed to the idea, discussion of results and writing of the paper. S.K. and M.D.  have performed simulations and S. K. developed the analytical treatment.

\begin{acknowledgments}
We acknowledge the  MULTIPLEX (No. 317532) EU project,
M.D. thanks Alan Danziger for first suggesting router software versions as a percolation problem. We also express gratitude to Shlomo Havlin, Damir Vuki\v{c}evi\'c, Marko Popovi\'c, Hrvoje \v{S}tefan\v{c}i\'c and Damir Koran\v{c}i\'{c} for helpful comments in the preparation of this manuscript.
\end{acknowledgments}

\onecolumngrid
\appendix
\section{Supplementary Information}
\newcommand{\tagS}{\tag{S\theequation}\stepcounter{equation}}
\section*{List of variables}
{\centering

\begin{tabular}{ c c }
 \hline
 \multicolumn{2}{ c } {Networks}\\
 \hline
 $N$ & Number of nodes \\
 $\bar k$ & Average degree \\
 $k_i$ & Degree of node $i$ \\
 $p_k$ & Degree distribution \\
 $\alpha$ & Exponent of scale free degree distribution \\
 $g_0$ & Generating function of degree \\
 $g_1$ & Generating function of excess degree \\
 \hline
 \multicolumn{2}{ c } {Colors}\\
 \hline
 $C$ & Number of colors \\
 $c\in 1,2,\dots C$ & A color \\
 $r_c$ & Color distribution \\
 \hline
 \multicolumn{2}{ c } {Standard percolation ingredients}\\
 \hline
 ${\mathcal L}$ & Set of nodes in the largest component (color blind) \\
 $u$ & Prob.\ of not being connected to giant comp.\ over a link \\
 $S$ & Size of giant component \\
 ${\mathcal L}_{\bar c}$ & Set of nodes in the largest component, after nodes of color c deleted\\
 $u_{\bar c}$ & Prob.\ of not being connected to giant ${\mathcal L}_{\bar c}$ over a link \\
 $S_{\bar c}$ & Size of giant ${\mathcal L}_{\bar c}$ \\
 \hline
 \multicolumn{2}{ c } {Percolation over color avoiding paths}\\
 \hline
 ${\mathcal L}_{\rm color}$ & Candidate set of nodes for the largest avoidable colors component\\
 $S_{\rm color}$ & Size of giant ${\mathcal L}_{\rm color}$ \\
 $B_{k,k'}$ & Prob.\ that out of $k$ links $k'$ connect to giant component \\
 $M_{k',\vec \kappa}$ & Prob.\ that out of $k'$ links $\kappa_1$ connect to color 1 etc. \\
 $P_{\vec \kappa}$ & Success probability having neighbors of colors acc. to $\vec \kappa$ \\
 $U_{\bar c}$ & Prob.\ that a link fails connecting to ${\mathcal L}_{\rm color}$ which already connects to ${\mathcal L}$ and 
 a node not having color $c$\\
 $S_{{\rm color},\infty}$ & Size of the set of all nodes being connected to giant component over two links or more \\
 \hline
 \hline
 $\beta$ & Critical exponent \\
 ${\bar k}_{\rm crit}$ & Critical value of average degree \\
\end{tabular}

}

\pagebreak

\section{Size of giant avoidable colors component in the configuration model}

We can find analytical results for $S_{\rm color}$ for random graph ensembles 
with randomly distributed colors in the limit of infinite graphs. 
These results can be used to estimate the situation in finite quenched networks. 
We are able to gain a general understanding including phase transitions. 
This knowledge can guide our understanding of real world networks. 

We use the generalized configuration model graph ensemble with $N$ nodes, 
where each degree sequences $\{k_i\}$ occurs with probability $\prod_i p_{k_i}$, 
with the degree distribution $p_{k}$. 
Additionally we want to assign to every node $i$ a color $c_i\in 1,2,\dots,C$. 
The color sequence $\{c_i\}$ has probability $\prod_i r_{c_i}$ with the color distribution $r_c$. 
For a graph $G_N$ out of the graph ensemble, 
$\mathcal{L}_{\rm color}$ has a certain size $N_{\rm color}(G_N)$. 
For the whole graph ensemble, we have to use the average value. 
By considering only giant contributions growing with network size, 
we have
\begin{align}
S_{\rm color} &= \lim_{N\to \infty}\sum_{G_N} P(G_N) \frac{N_{\rm color}(G_N)}{N},
\tagS
\end{align}
where $P(G_N)=\prod_i p_{k_i} \omega \prod_i r_{c_i}$ is the probability 
to have the graph $G_N$ of size $N$, including $\omega$, 
the probability of the connection scheme of $G_N$ as a matching of half edges. 

\subsection{On the construction and maximality of $\mathcal{L}_{\rm color}$}
By construction, every node pair in $\mathcal{L}_{\rm color}$ is color-avoiding connected. 
Furthermore, $\mathcal{L}_{\rm color}$ is maximal and therefore it is a color-avoiding connected component, if it includes for every color $c$ at least one node out of $\mathcal{L}_{\bar c}$.
If for every color $c$, $\mathcal{L}_{\rm color}$ includes at least one node out of $\mathcal{L}_{\bar c}$, 
it is maximal and therefore it is an avoidable colors component. 
To prove this, assume it was not maximal. 
Then a node can be added which is (a) connected to every node in $\mathcal{L}_{\rm color}$ and (b) excluded 
from $\mathcal{L}_{\rm color}$;  it does not connect to $\mathcal{L}_{\bar c'}$ for some color $c'$. 
Consequently, it cannot connect to the nodes in 
$\mathcal{L}_{\rm color}$ which belong to $\mathcal{L}_{\bar c'}$, which contradicts (a). 

\begin{figure}
\centering
  \begin{minipage}{0.18\columnwidth}\centering
    (a)\\
    \includegraphics[width=1.0\columnwidth]{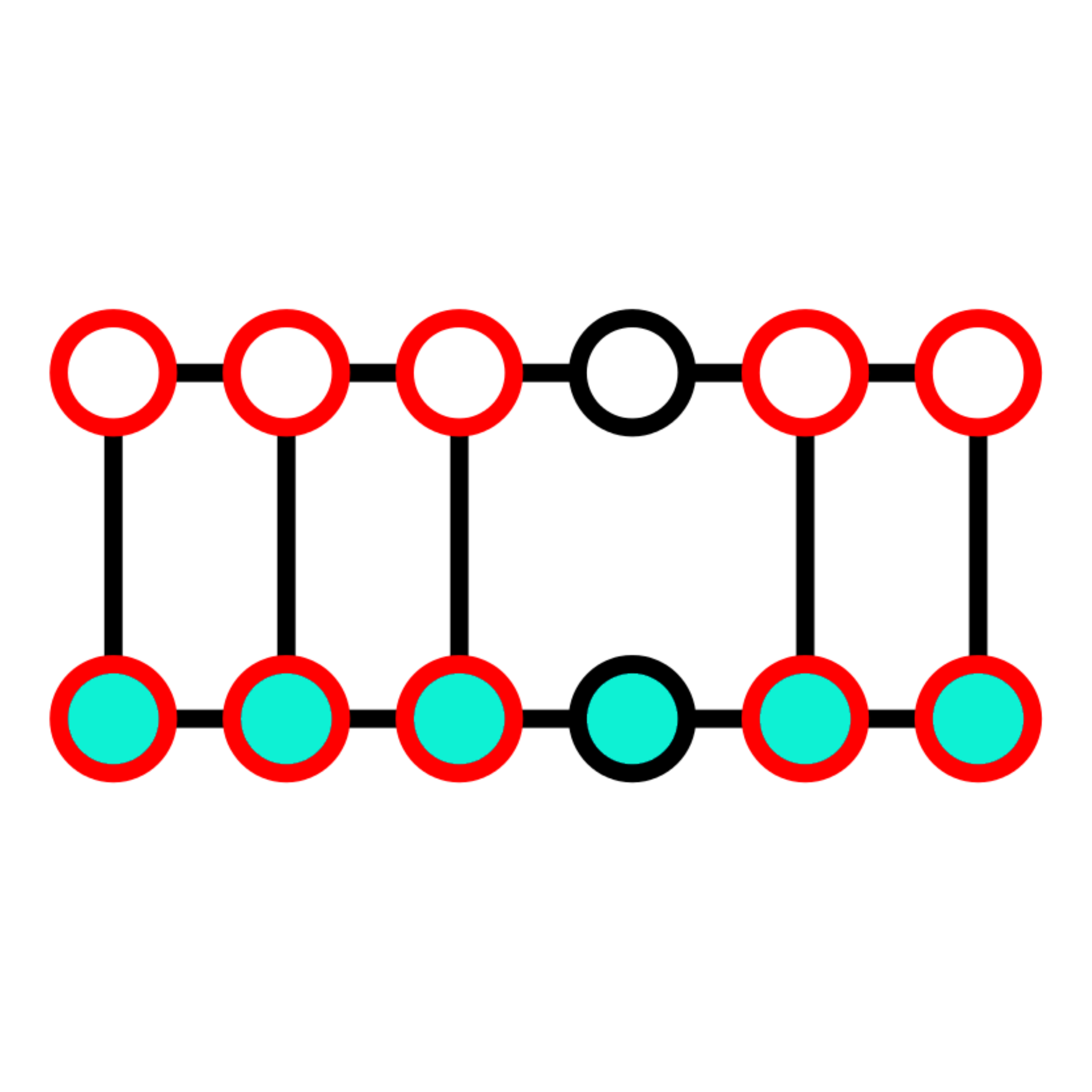}
  \end{minipage}
  \begin{minipage}{0.18\columnwidth}\centering
    (b)\\
    \includegraphics[width=1.0\columnwidth]{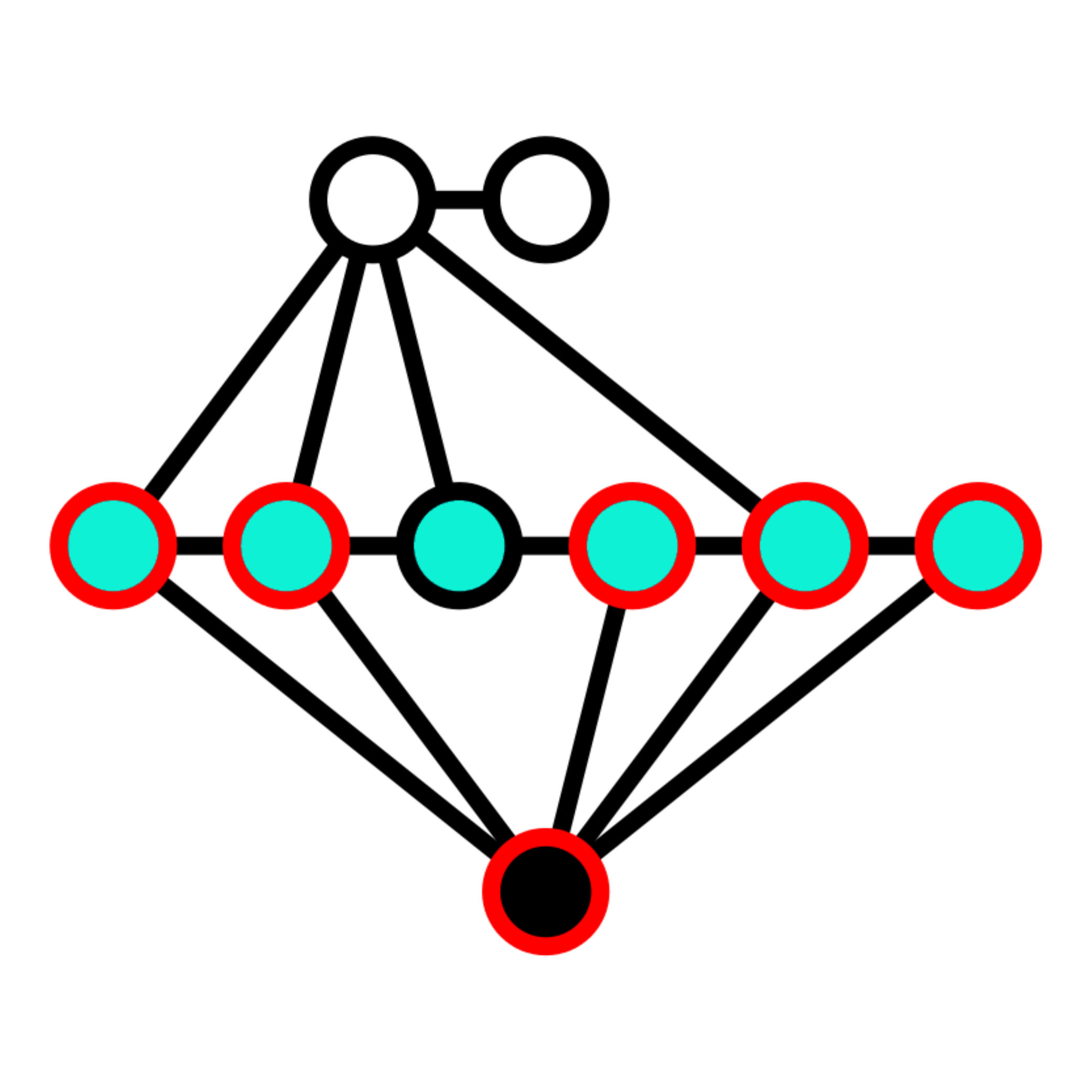}
  \end{minipage}
  \begin{minipage}{0.18\columnwidth}\centering
    (c)\\
    \includegraphics[width=1.0\columnwidth]{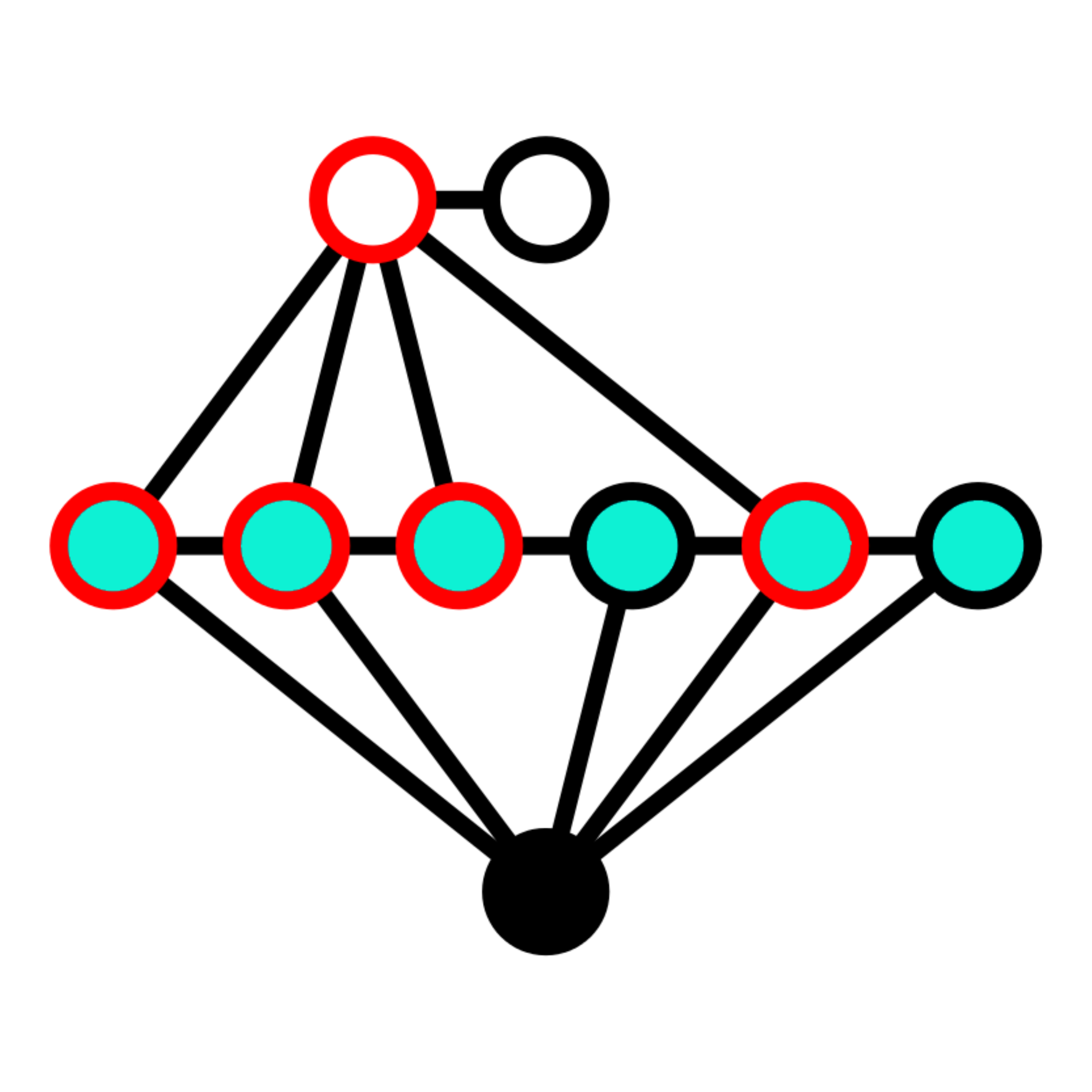}
  \end{minipage}
  \begin{minipage}{0.18\columnwidth}\centering
    (d)\\
    \includegraphics[width=1.0\columnwidth]{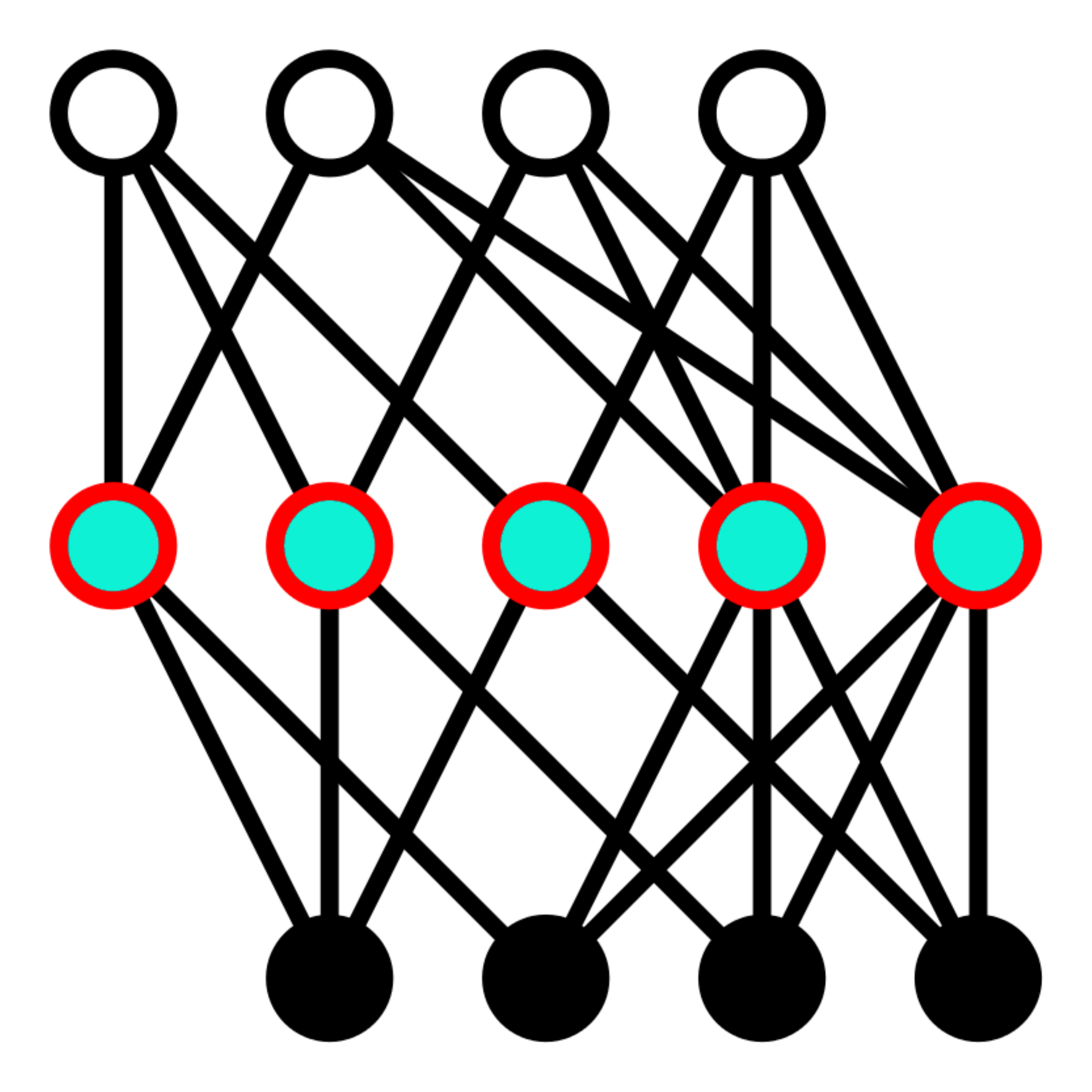}
  \end{minipage}
  \begin{minipage}{0.18\columnwidth}\centering
    (e)\\
    \includegraphics[width=1.0\columnwidth]{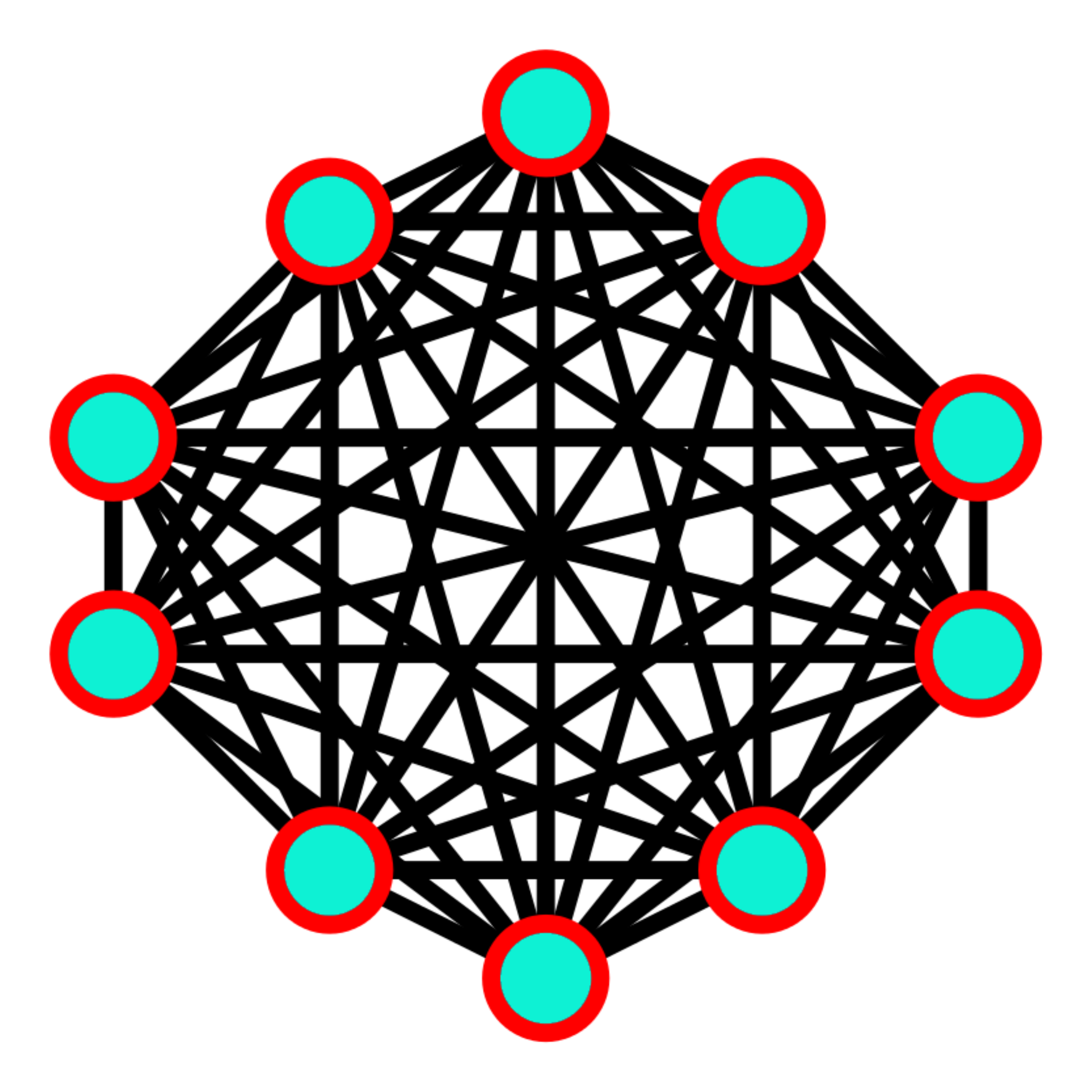}
  \end{minipage}
\caption{ Color avoiding components may overlap, as shown in \textbf{(b)} and \textbf{(c)}. 
    Color avoiding components can assume diverse forms. 
    In a chain \textbf{(a)}, paths between nodes of one color exist and can be 
    reached by connections between nodes of different colors. In \textbf{(b)}, the black 
    node serves as an alternative path provider for the blue nodes. 
    The graph \textbf{(d)} does not need any connection among nodes of the same color, 
    but there is a massive overhead of nodes and connections to achieve 
    color-avoiding connectivity of the blue nodes. A clique is a color avoiding component \textbf{(e)}.}
\label{fig:my_label}
\end{figure}

\subsection{Question and connection to percolation theory}

For calculating $S_{\rm color}$ in the random graph ensemble, 
we will follow ideas of Erd\H{o}s and R\'{e}nyi~\cite{erd-1959random} and Newman~\cite{newman-2001random}. 
For calculating the size of the giant component, 
they used probabilities of connections for a single node in the graph ensemble. 
As we have to extend the method to a gradual procedure with conditional probabilities, 
it is useful to introduce the original method in detail with a shifted viewpoint. 

\begin{figure}[htb]
  \begin{minipage}[b]{0.18\linewidth}
    \begin{center}
      \includegraphics[width=0.99\columnwidth]{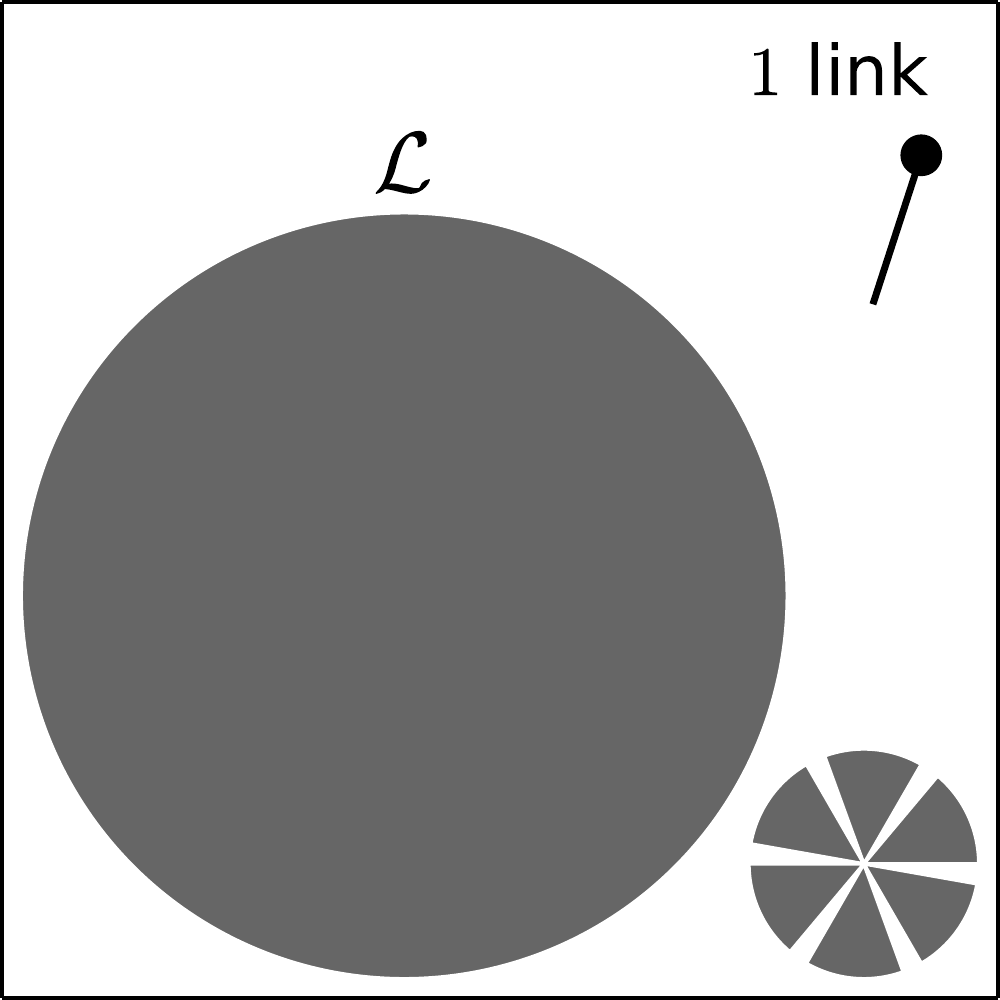}
   \end{center}
  \end{minipage}
  \begin{minipage}[b]{0.05\linewidth}
    \begin{center}
      {\large $\xrightarrow{u}$}\\
      \vspace{15mm}
    \end{center}
  \end{minipage}
  \begin{minipage}[b]{0.18\linewidth}
    \begin{center}
      \includegraphics[width=0.99\columnwidth]{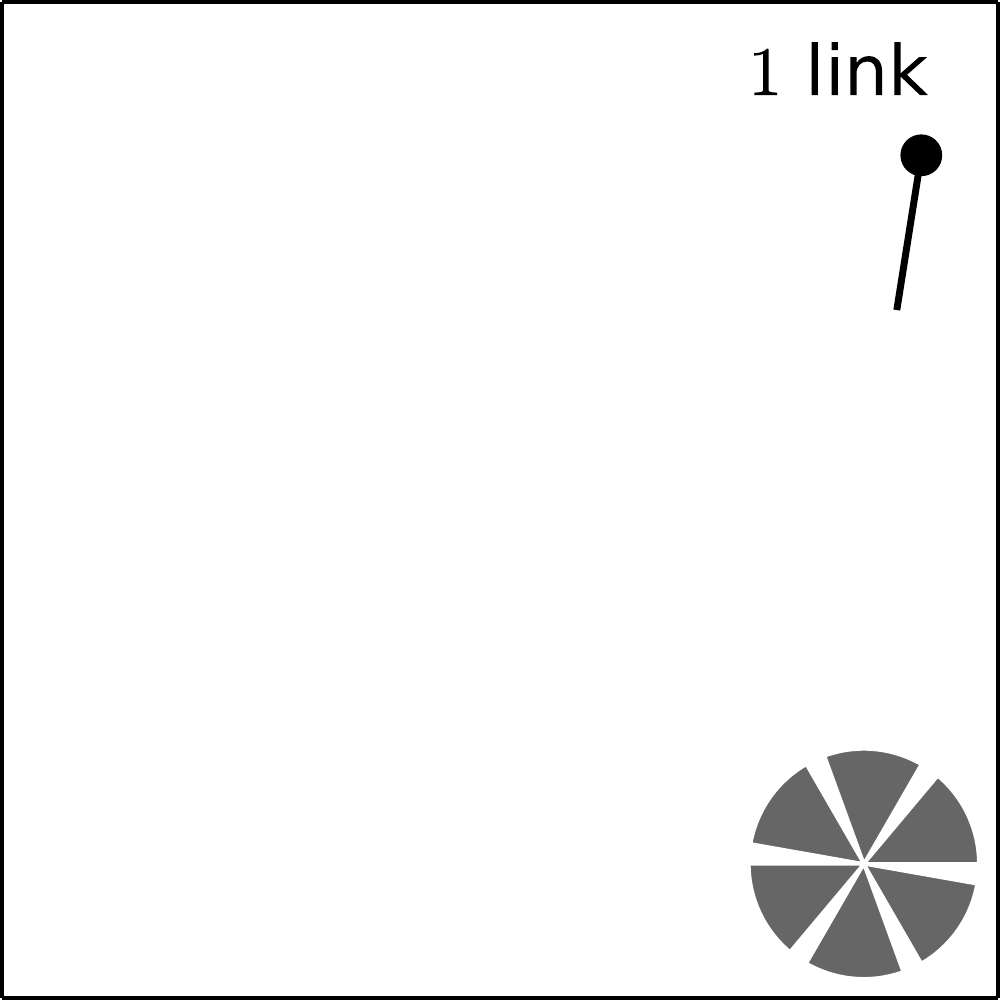}
    \end{center}
  \end{minipage}
  \begin{minipage}[b]{0.05\linewidth}
  \ 
  \end{minipage}
  \begin{minipage}[b]{0.18\linewidth}
    \begin{center}
      \includegraphics[width=0.99\columnwidth]{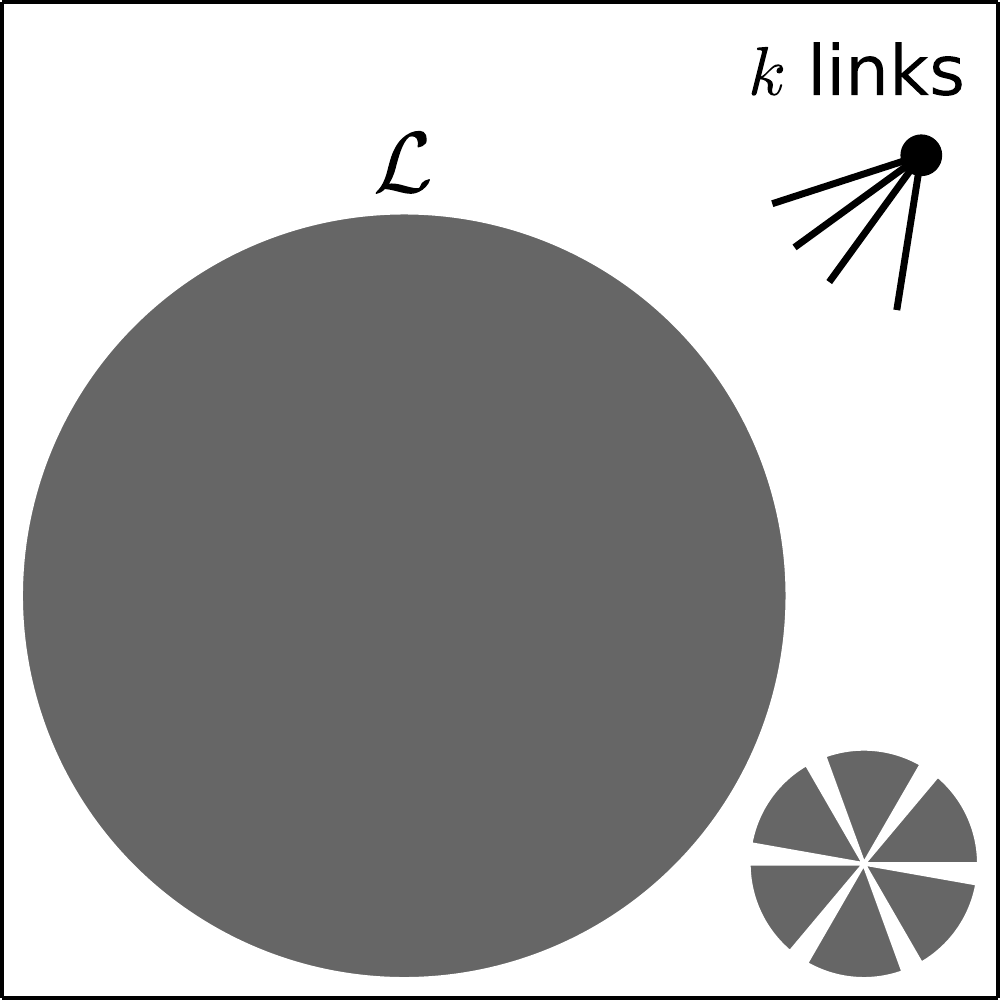}
    \end{center}
  \end{minipage}
  \begin{minipage}[b]{0.07\linewidth}
    \begin{center}
      {\large $\xrightarrow{1-u^k}$}\\
      \vspace{15mm}
    \end{center}
  \end{minipage}
  \begin{minipage}[b]{0.18\linewidth}
    \begin{center}
      \includegraphics[width=0.99\columnwidth]{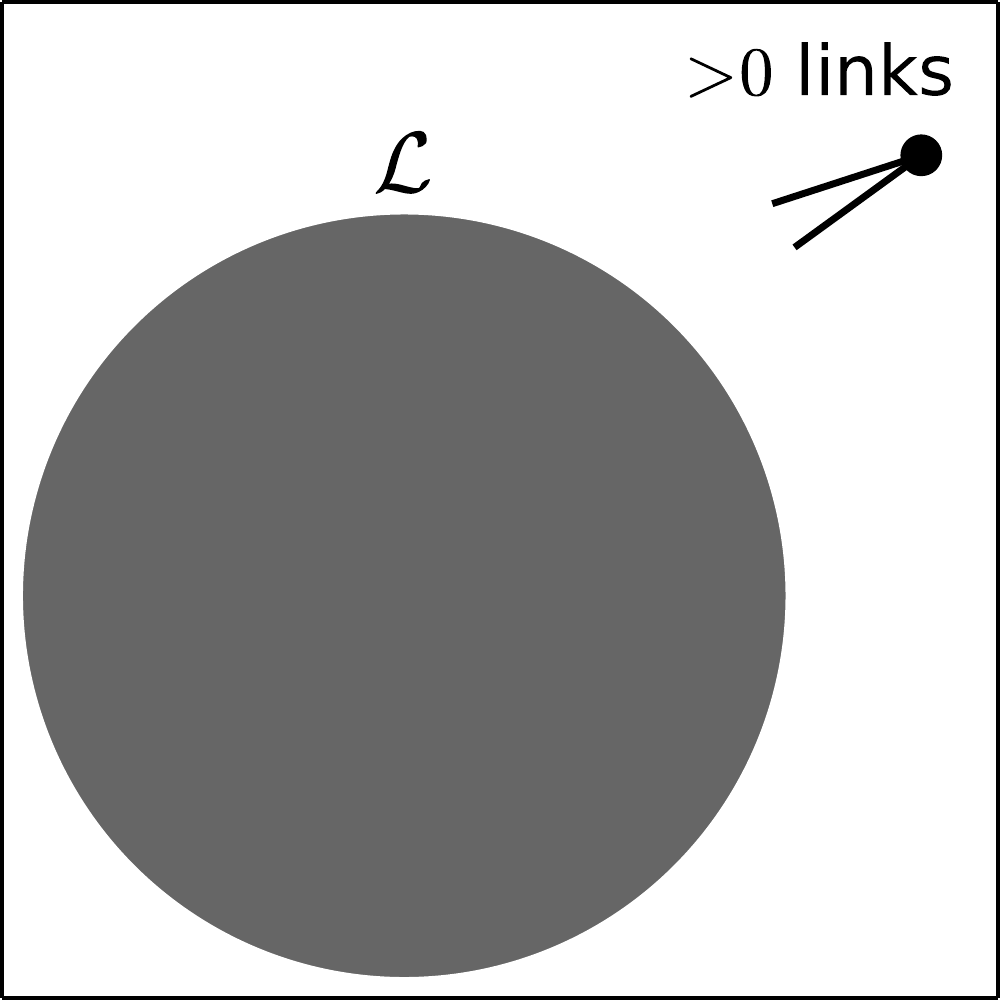}
    \end{center}
  \end{minipage}
    \caption{We base our theory on the method to calculate the size of 
    normal giant components, as illustrated in this figure. Using a self 
    consistency equation, the probability $u$ can be calculated. This is the 
    probability, that a node is not connected to the giant component over a 
    single link (see on the left). On the right, the probability for a node 
    with $k$ links is illustrated to have at least one link connecting to 
    the giant component. $u^k$ is the probability that all links fail.}
    \label{fig:gc}
\end{figure}
Lets denote with $\mathcal{L}$ the set of all nodes belonging to the largest component. 
In figure~\ref{fig:gc} on the outer left, a possible situation is illustrated. 
The largest component contains of a large part of the network, 
and the remaining nodes belong to smaller components. 
We have to calculate the size $S$ of the giant component, 
meaning the average relative size of $\mathcal{L}$ in the network ensemble 
in the limit of infinite network size. 
For this we can define the average probability $u$ 
that a node fails to connect to $\mathcal{L}$ over one particular link.  
This is illustrated in the figure with the left part. 
Again, the thermodynamic limit $N\to \infty$ is implied. 
With the definition of $u$ at hand, we can calculate $S$ in two steps: 
First, using a self consistency equation, $u$ is calculated. 
The probability $u$ is identical to the probability 
that the neighbor does not connect to the giant component over any of the remaining links, 
\begin{align}
u &= g_1(u),\quad g_1(z)=\sum_k q_k z^k.\label{eq:u}
\tagS
\end{align}
In this equation, $g_1$ is the generating function 
of excess degree $q_k=(k+1)p_{k+1}/\bar{k}$. 
For important degree distributions as e.g. Poisson or scale-free, 
the equation for $u$ can only be solved numerically. 
The second step is an averaging over nodes with various degrees $k$. 
The probability to connect to the giant component over any of $k$ links is $(1-u^k)$, 
meaning that not all links fail at the same time. 
This is illustrated in the figure on the right. 
As a node which connects to the giant component belongs to it, 
\begin{align}
S &= \sum_{k=0}^{\infty}p_k (1-u^k) = 1-g_0(u),\quad g_0(z)=\sum_k p_k z^k.\label{eq:S}
\tagS
\end{align}

\begin{figure}[htb]
  \begin{minipage}[b]{0.245\linewidth}
    \begin{center}
    \includegraphics[width=0.99\columnwidth]{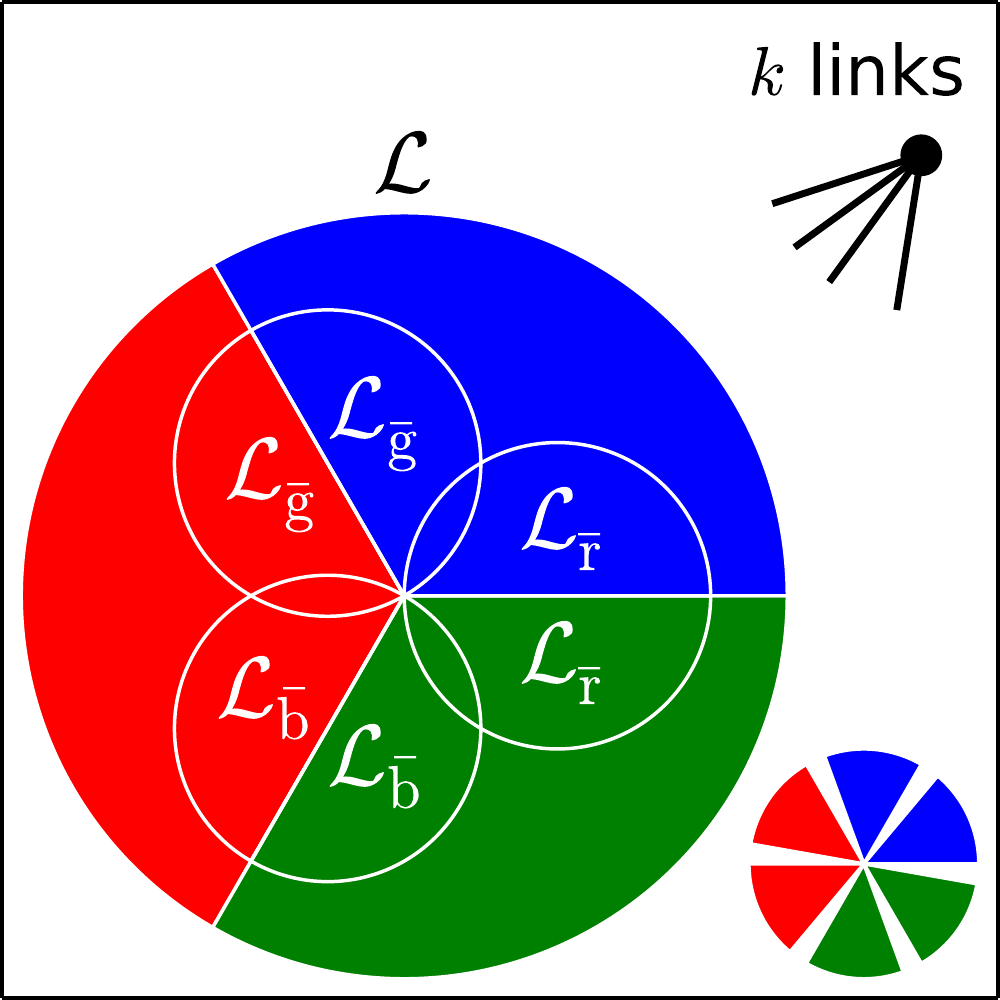}
   \end{center}
  \end{minipage}
  \begin{minipage}[b]{0.1\linewidth}
    \begin{center}
      {\Large $\xrightarrow{?}$}\\
      \vspace{20mm}
    \end{center}
  \end{minipage}
  \begin{minipage}[b]{0.6\linewidth}
    \begin{center}
    \includegraphics[height=0.4\columnwidth]{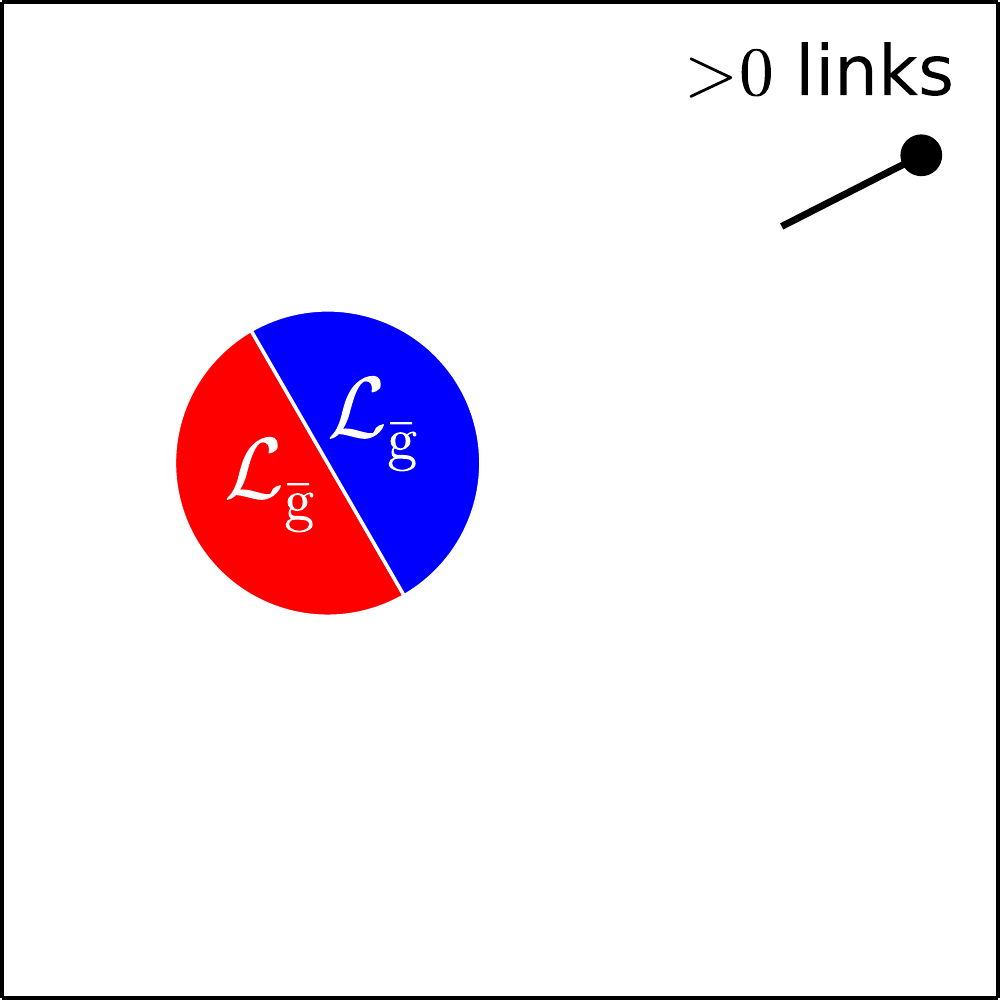}
     \hspace{-1mm}
    \includegraphics[trim=100 0 0 0,clip,height=0.4\columnwidth]{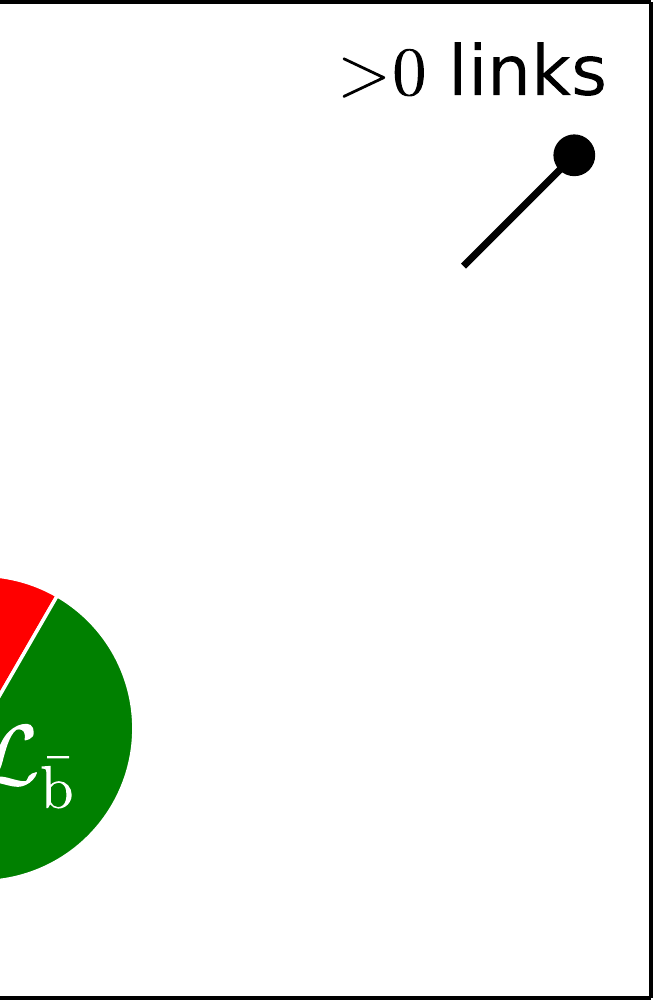}
     \hspace{-1mm}
    \includegraphics[trim=100 0 0 0,clip,height=0.4\columnwidth]{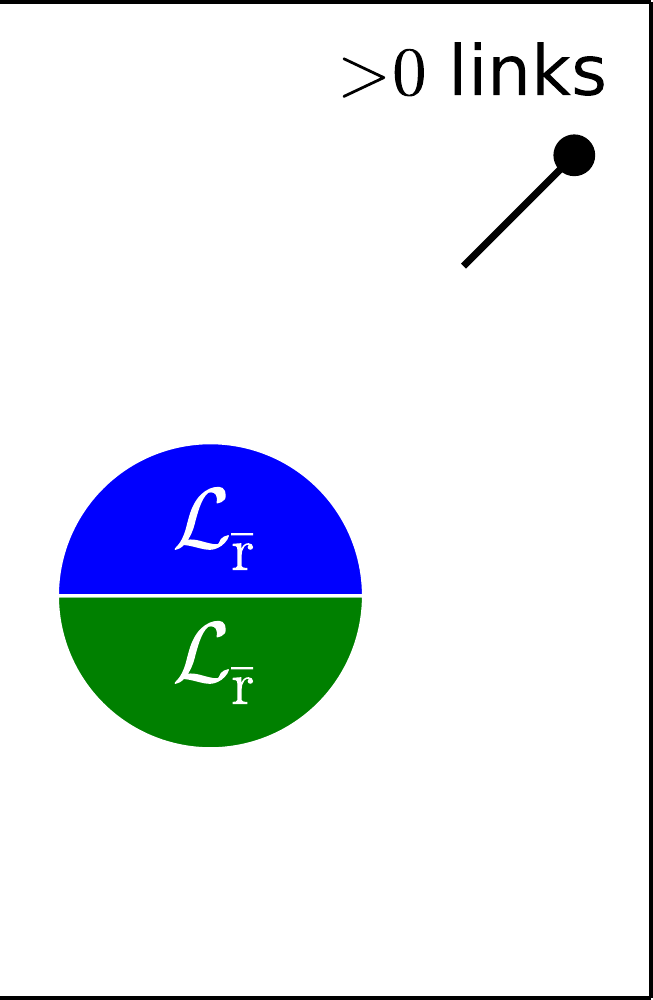}
   \end{center}
  \end{minipage}
    \caption{We have to calculate the probability, if a node with $k$ links is 
    for every color $c$ connected to the giant component $\mathcal{L}_{\bar c}$ with deleted 
    color $c$. All connections over at least one link have to exist at the same 
    time. We illustrate this question with the three colors red ($c=$r), green 
    ($c=$g) and blue ($c=$b). If a link connects to $\mathcal{L}_{\bar{\rm g}}$, it for 
    sure does not connect to $\mathcal{L}_{\bar c}$ for one of the other colors. This kind of 
    dependence forces us to use a stepwise calculation with conditional probabilities.}
    \label{fig:question}
\end{figure}

In analogy to the procedure described above, 
we will calculate $S_{\rm color}$ as the probability 
that a randomly chosen node belongs to $\mathcal{L}_{\rm color}$. 
This has to be evaluated in the graph ensemble of infinite size. 
As we will perform an averaging over nodes with various degrees $k$, 
the following question has to be answered: 
What is the probability that a node with $k$ links connects 
to a giant $\mathcal{L}_{\bar c}$ for all colors $c$ at the same time. 
This is illustrated in figure~\ref{fig:question}. 
On the left, the situation for a graph with colors on the nodes is illustrated. 
Nodes of all colors might be in the largest component. 
After deleting all nodes of one color $c$, 
the remaining largest component $\mathcal{L}_{\bar c}$ 
might still contain a large part of all nodes in $\mathcal{L}$. 
The condition for the node belonging to $\mathcal{L}_{\rm color}$ 
is illustrated on the right of the figure. 

We will use the same two steps to attack this problem, 
as described for calculating the giant component above. 
First, we provide some single link probabilities 
which can be used as primitives for the further calculations. 
Second, we combine the single link probabilities to calculate $S_{\rm color}$.

\subsection{Single link probabilities}

\begin{figure}[htb]
   \begin{center}
    \includegraphics[width=0.2\columnwidth]{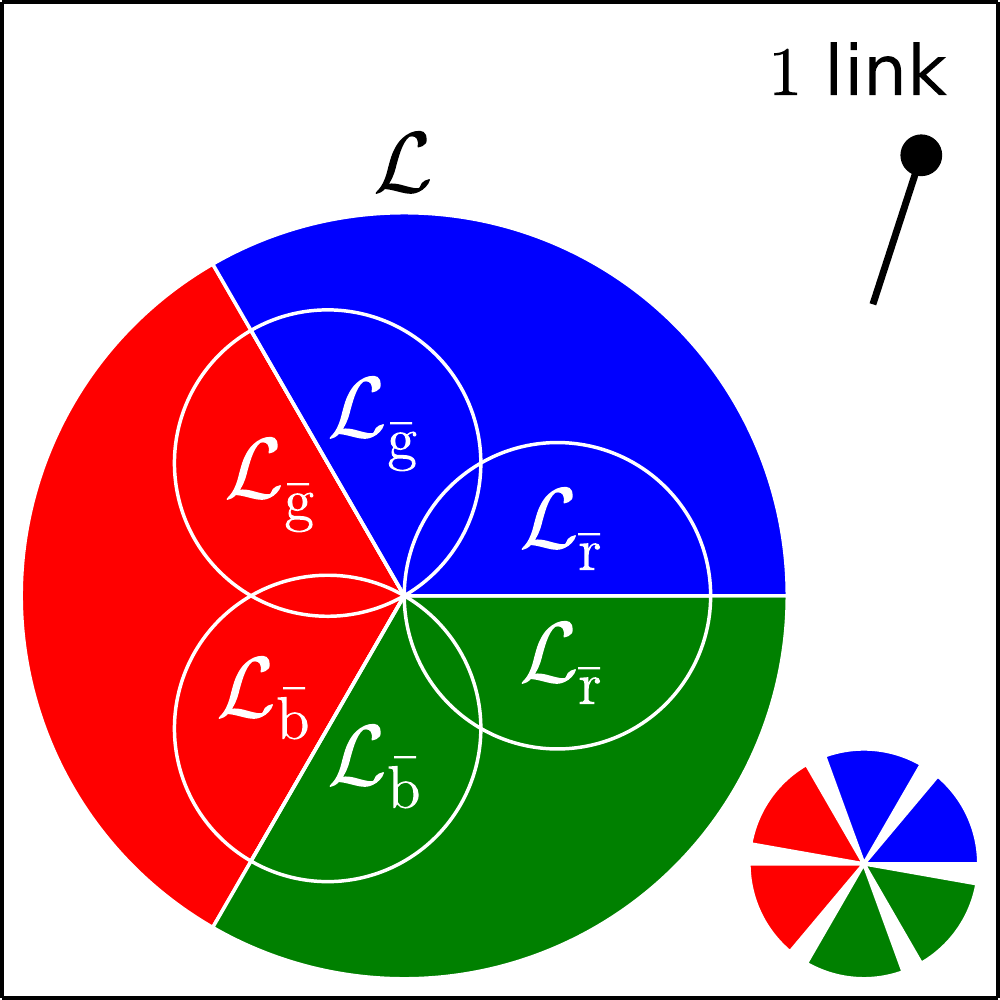}\qquad
    \includegraphics[width=0.2\columnwidth]{sets_1_all.pdf}\qquad
    \includegraphics[width=0.2\columnwidth]{sets_1_all.pdf}\qquad
    \includegraphics[width=0.2\columnwidth]{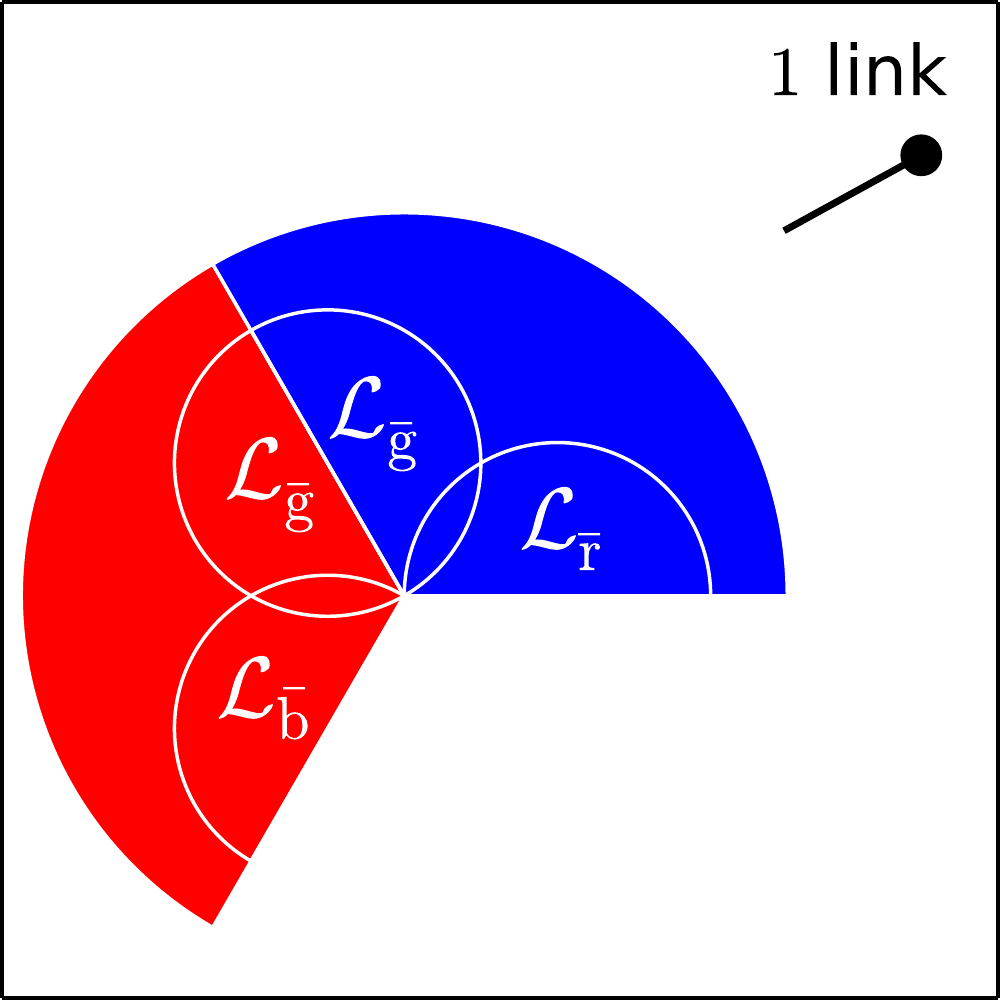}\\
     \vspace{1mm}
      { \hspace{4mm} $\downarrow u$\hspace{36mm}$\downarrow r_{\rm g}$ 
      \hspace{36mm}$\downarrow u_{\bar{\rm g}}$ \hspace{36mm}$\downarrow U_{\bar{\rm g}}$}\\
     \vspace{1mm}
    \includegraphics[width=0.2\columnwidth]{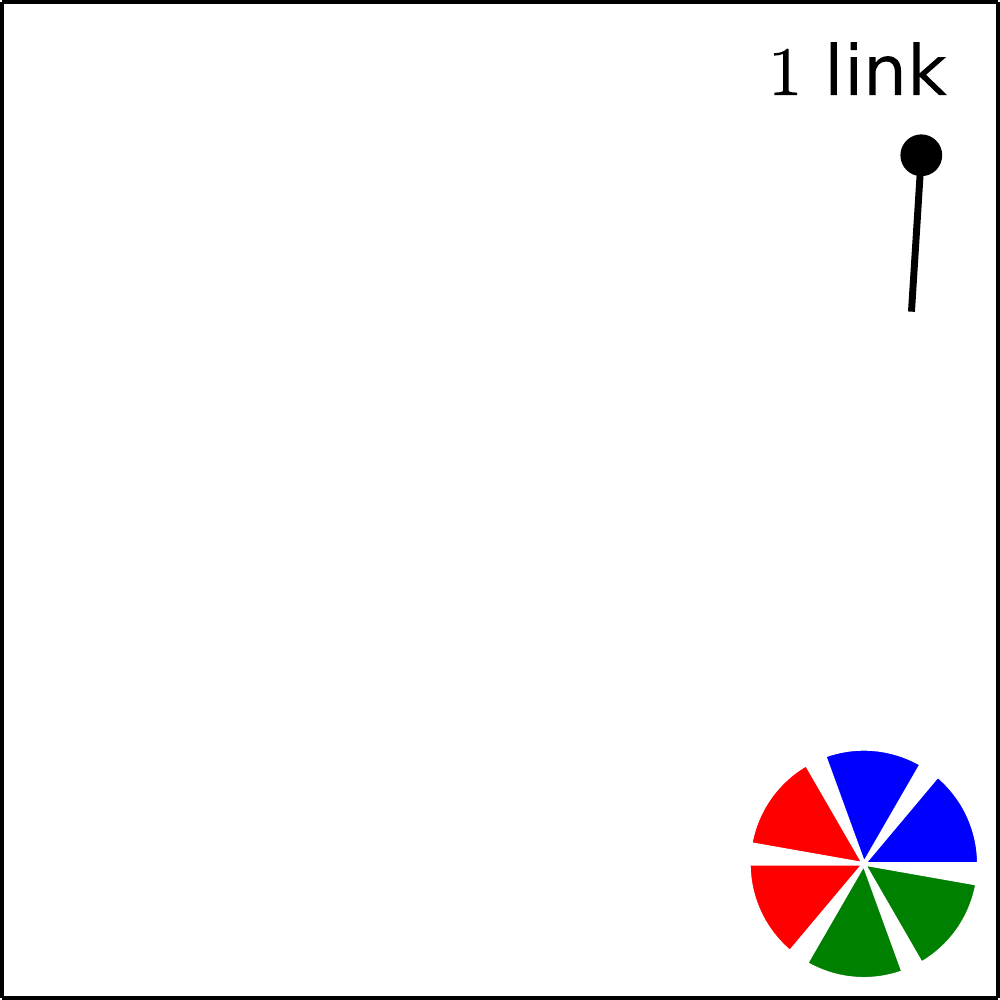}\qquad
    \includegraphics[width=0.2\columnwidth]{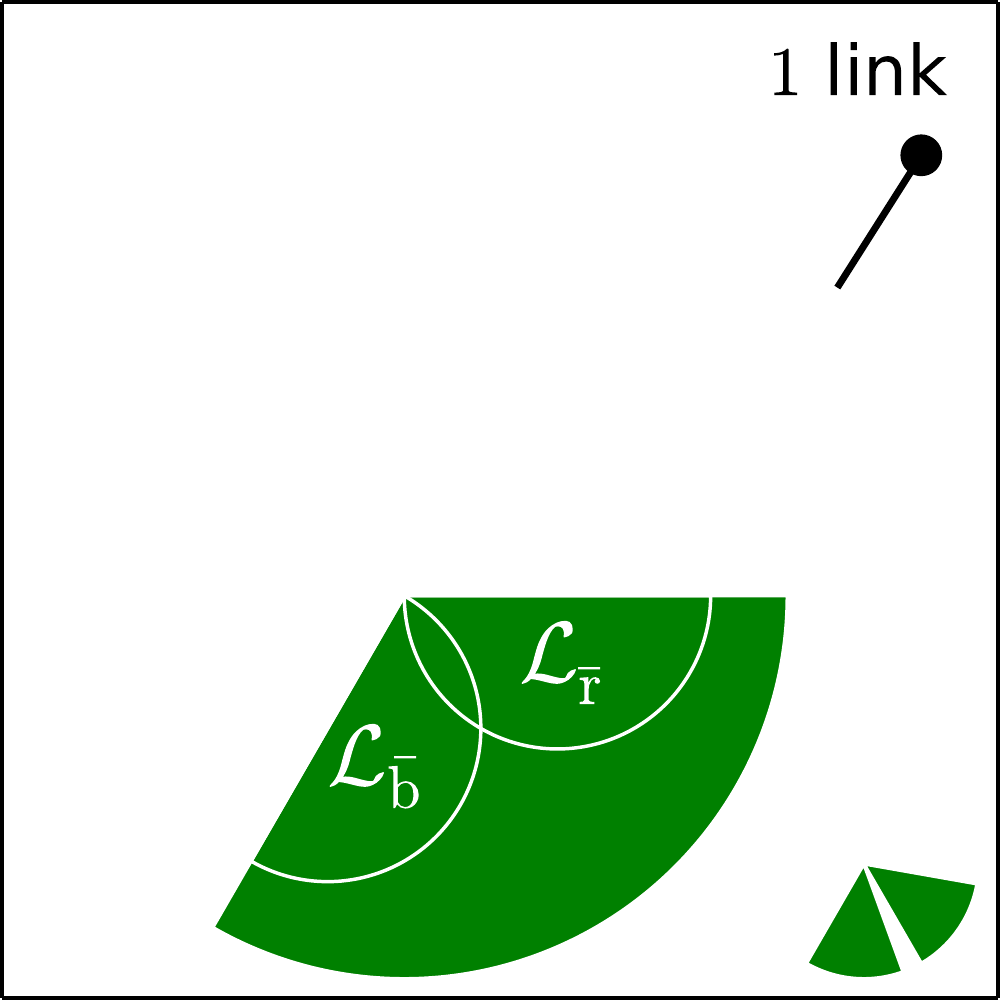}\qquad
    \includegraphics[width=0.2\columnwidth]{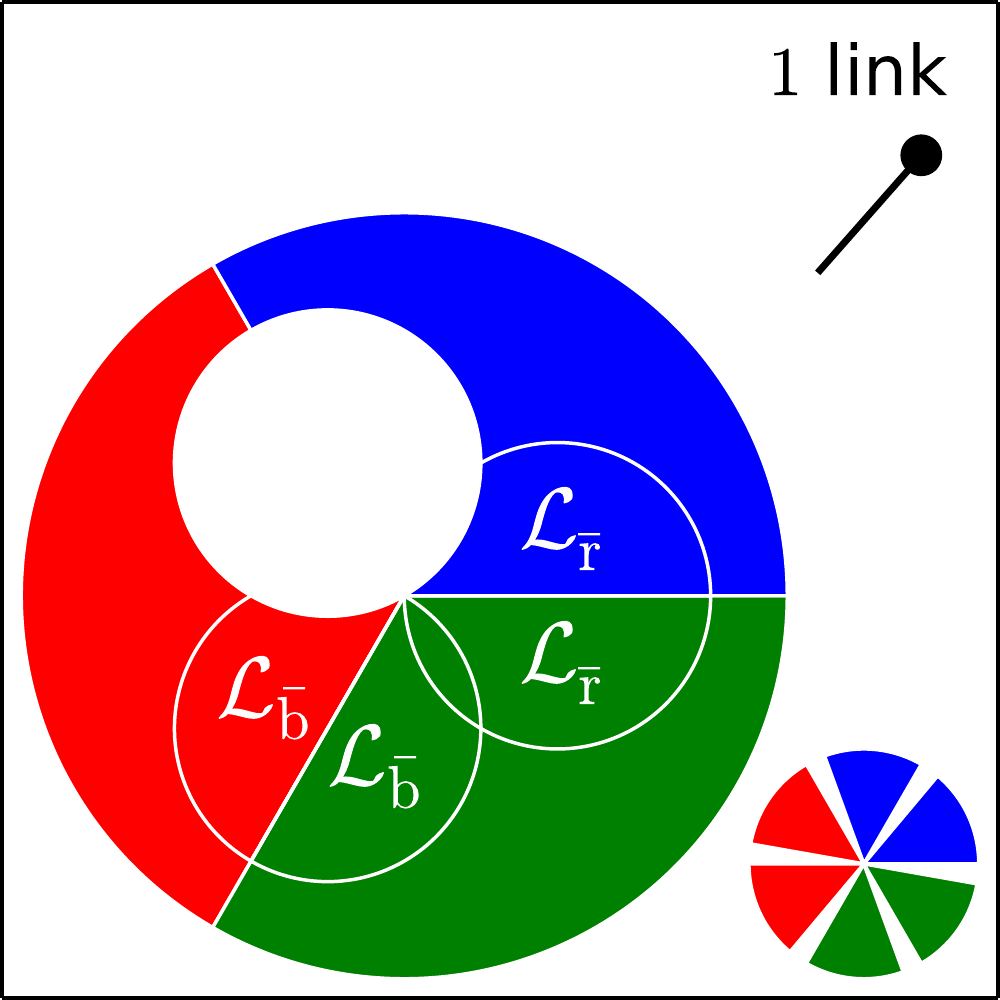}\qquad
    \includegraphics[width=0.2\columnwidth]{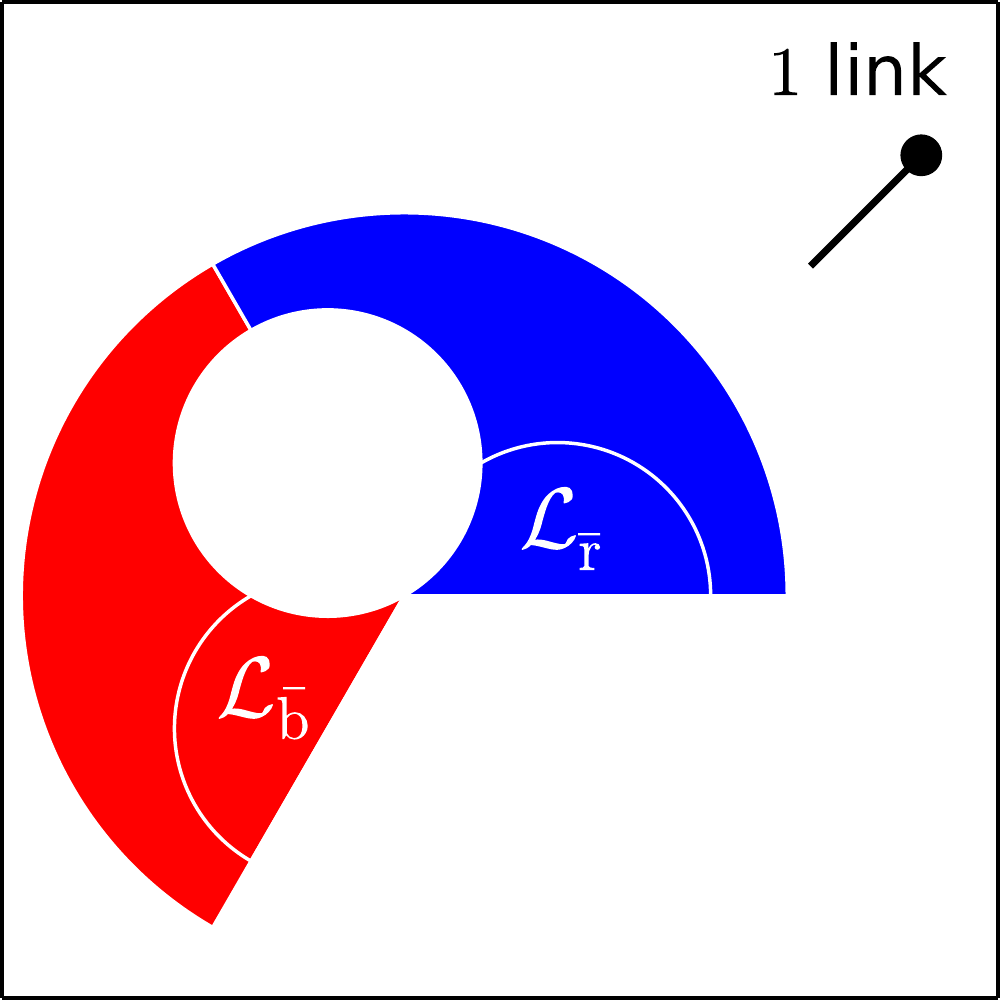}
   \end{center}
  \caption{Probabilities for a single link to connect to different parts of the network. We 
  use these probabilities as primitives to calculate the probability for many links. While 
  $u$, $r_c$ and $u_{\bar c}$ can be calculated with standard methods invented for the 
  configuration model before, the conditional probability $U_{\bar c}$ can be calculated as 
  a combination of the others.}
  \label{fig:primitives}
\end{figure}
We already gave equation~\ref{eq:u} for calculating the probability $u$. 
In the case of colors on the nodes, as illustrated 
in figure~\ref{fig:primitives} on the left, the colors can simply be ignored. 
We further need the probability to connect to a node of color $c$ which is simply $r_c$. 
This is illustrated in the second column of the figure. 
We further introduce $u_{\bar c}$, 
the probability that a single link does not connect to a giant $\mathcal{L}_{\bar c}$. 
See the third column of the figure for an illustration. 
This can be calculated using percolation theory 
for random attack by solving 
\begin{align}
u_{\bar c} &= r_c + (1-r_c) g_1(u_{\bar c}).\label{eq:u_c}
\tagS
\end{align}

\begin{figure}[htb]
  \begin{minipage}[b]{0.2\linewidth}
    \begin{center}
    \includegraphics[width=0.99\columnwidth]{sets_1_all.pdf}\\
      \vspace{42mm}
   \end{center}
  \end{minipage}
  \begin{minipage}[b]{0.15\linewidth}
    \begin{center}
      $(1-u)(1-r_{\rm g})$\\
      {\large $\rightarrow$\\}
      \vspace{20mm}
      $1-u_{\bar{\rm g}}$\\
      {\large $\searrow$\\}
      \vspace{27mm}
    \end{center}
  \end{minipage}
  \begin{minipage}[b]{0.2\linewidth}
    \begin{center}
    \includegraphics[width=0.99\columnwidth]{sets_1_no_2_gc.pdf}\\
     \vspace{1mm}
    {\large $\downarrow$} $1-U_{\bar{\rm g}}$\\
     \vspace{1mm}
    \includegraphics[width=0.99\columnwidth]{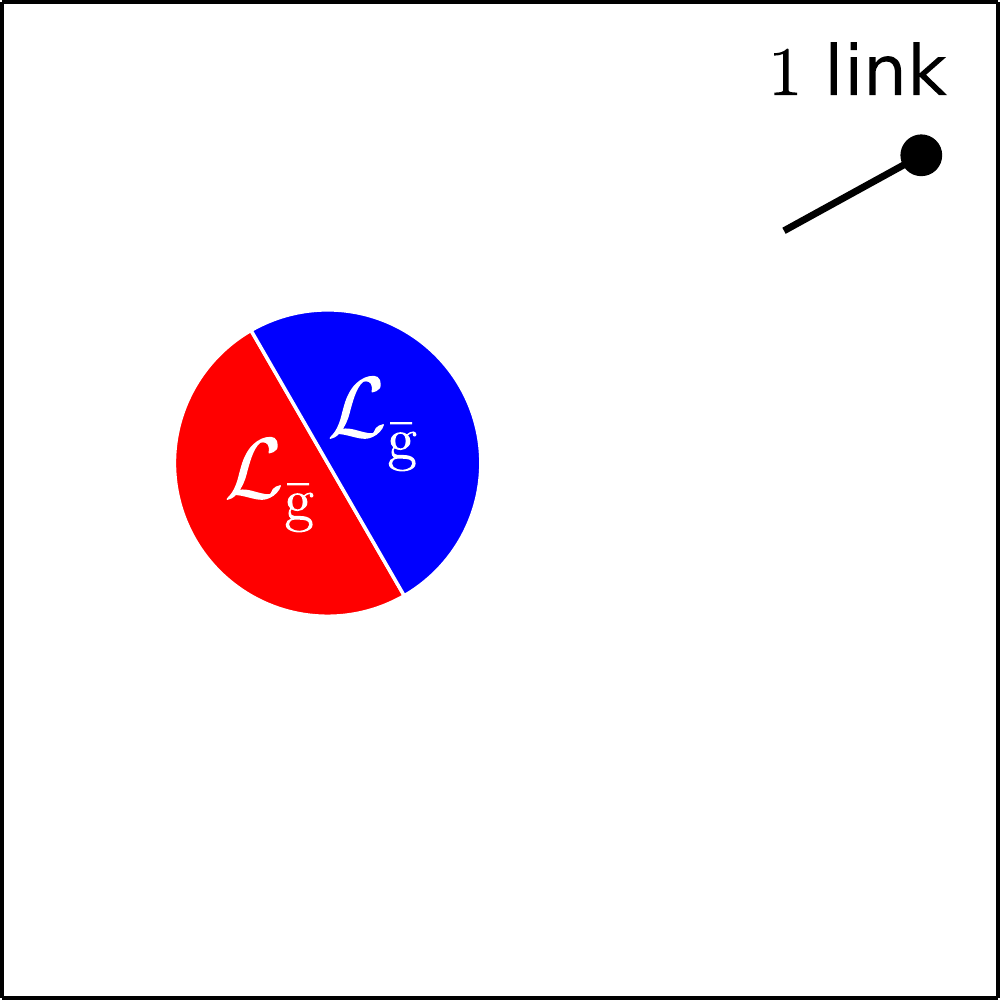}\\
   \end{center}
  \end{minipage}
    \caption{This figure illustrates the calculation of $U_{\bar{\rm g}}$ 
    using the equality $(1-u)(1-r_{\rm g})(1-U_{\bar{\rm g}})=1-u_{\bar{\rm g}}$. For 
    that, we have assumed independence of the qualities of the link under 
    consideration, especially of the color it connects to and if it connects to 
    the giant component. }
    \label{fig:U_c}
\end{figure}
Unfortunately, $u_{\bar c}$ cannot be used directly for calculating $S_{\rm color}$. 
If we look at the same link, 
the probabilities $u_{\bar c}$ are dependent for different colors. 
The most obvious argument is that always $\Pi_c (1-u_{\bar c})=0$, 
as a link must at least miss one of the $\mathcal{L}_{\bar c}$. 
Instead, we will use the conditional probability $U_{\bar c}$, 
as illustrated with the outer right column of the figure. 
The precondition is that a link connects to the giant component 
and the node it connects to has not color $c$. 
$U_{\bar c}$ is the probability that such a link connects to $\mathcal{L}_{\bar c}$. 
For calculating it, we use the primitives introduced so far, 
as illustrated in figure~\ref{fig:U_c}. 
Assuming independence of the probabilities $(1-u)$ for connecting to the giant component 
and $(1-r_c)$ for not connecting to a node of color $c$, 
the precondition of $U_{\bar c}$ can be constructed. 
In this way, we can construct $(1-u_{\bar c})$ using the probability we are searching for: 
$(1-u_{\bar c}) = (1-u)(1-r_c)(1-U_{\bar c})$. With this we find  
\begin{align}
U_{\bar c} &= 1 - \frac{1-u_{\bar c}}{(1-u)(1-r_c)}.\label{eq:U_c}
\tagS
\end{align}
If $(1-u)(1-r_c)=0$, the precondition holds for an empty set of nodes. 
In this case we define $U_{\bar c}=1$.
Notice that the additional information of the explicit color, instead of only stating that the color 
is not $c$, does not alter the results, as a further restriction of the colors 
would meat the numerator and denominator identically and therefore would cancel out.

\subsection{Averaging over link distributions}

\begin{figure}[htb]
  \begin{minipage}[b]{0.22\linewidth}
    \begin{center}
    \includegraphics[width=0.99\columnwidth]{sets_k_all.pdf}
   \end{center}
  \end{minipage}
  \begin{minipage}[b]{0.1\linewidth}
    \ 
  \end{minipage}
  \begin{minipage}[b]{0.54\linewidth}
    \begin{center}
    \includegraphics[height=0.4\columnwidth]{sets_k_gc_no_2.pdf}
     \hspace{-1mm}
    \includegraphics[trim=100 0 0 0,clip,height=0.4\columnwidth]{sets_k_gc_no_3.pdf}
     \hspace{-1mm}
    \includegraphics[trim=100 0 0 0,clip,height=0.4\columnwidth]{sets_k_gc_no_1.pdf}
   \end{center}
  \end{minipage}\\
     \vspace{2mm}
      {\large \hspace{-30mm} $\downarrow B_{k,k'}$\hspace{90mm}$\uparrow P_{\vec \kappa}$ }\\
     \vspace{2mm}
  \begin{minipage}[b]{0.22\linewidth}
    \begin{center}
    \includegraphics[width=0.99\columnwidth]{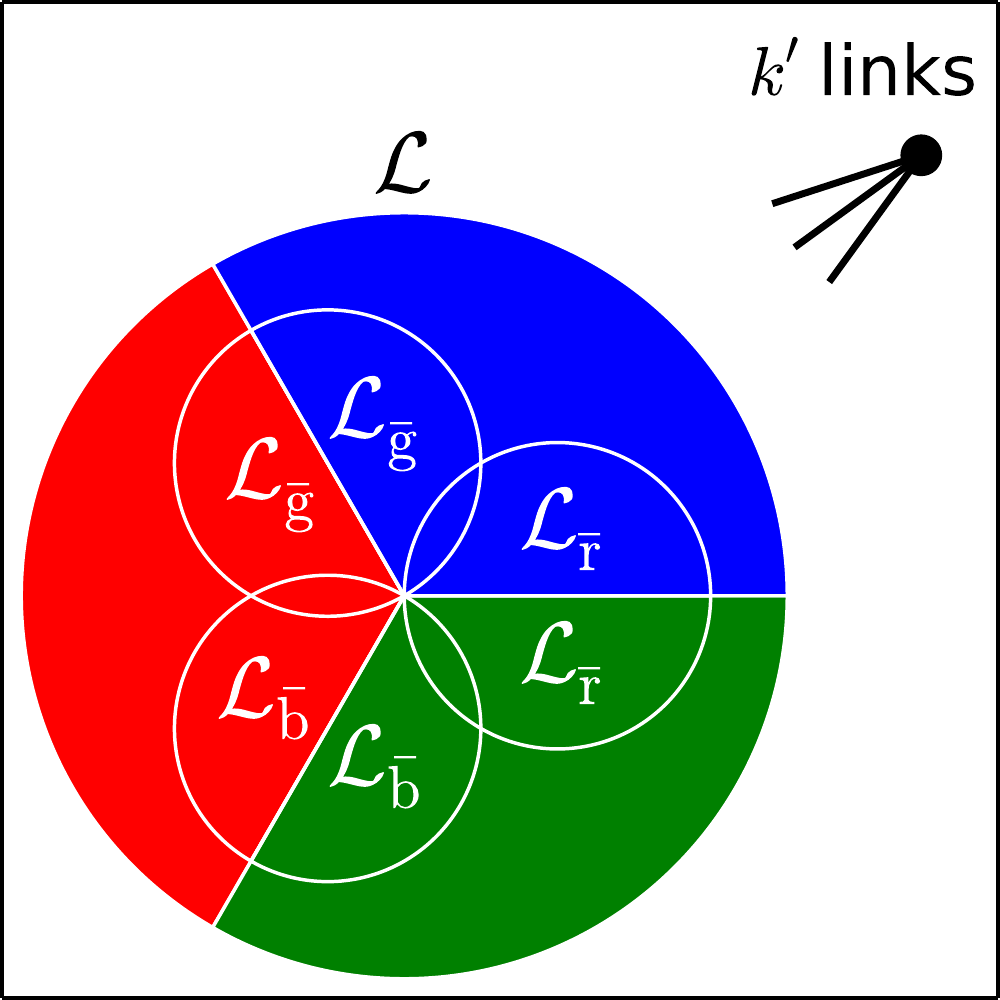}
   \end{center}
  \end{minipage}
  \begin{minipage}[b]{0.1\linewidth}
    \begin{center}
      {\Large $\xrightarrow{M_{k',\vec{\kappa}}}$}\\
      \vspace{20mm}
    \end{center}
  \end{minipage}
  \begin{minipage}[b]{0.54\linewidth}
    \begin{center}
    \includegraphics[height=0.4\columnwidth]{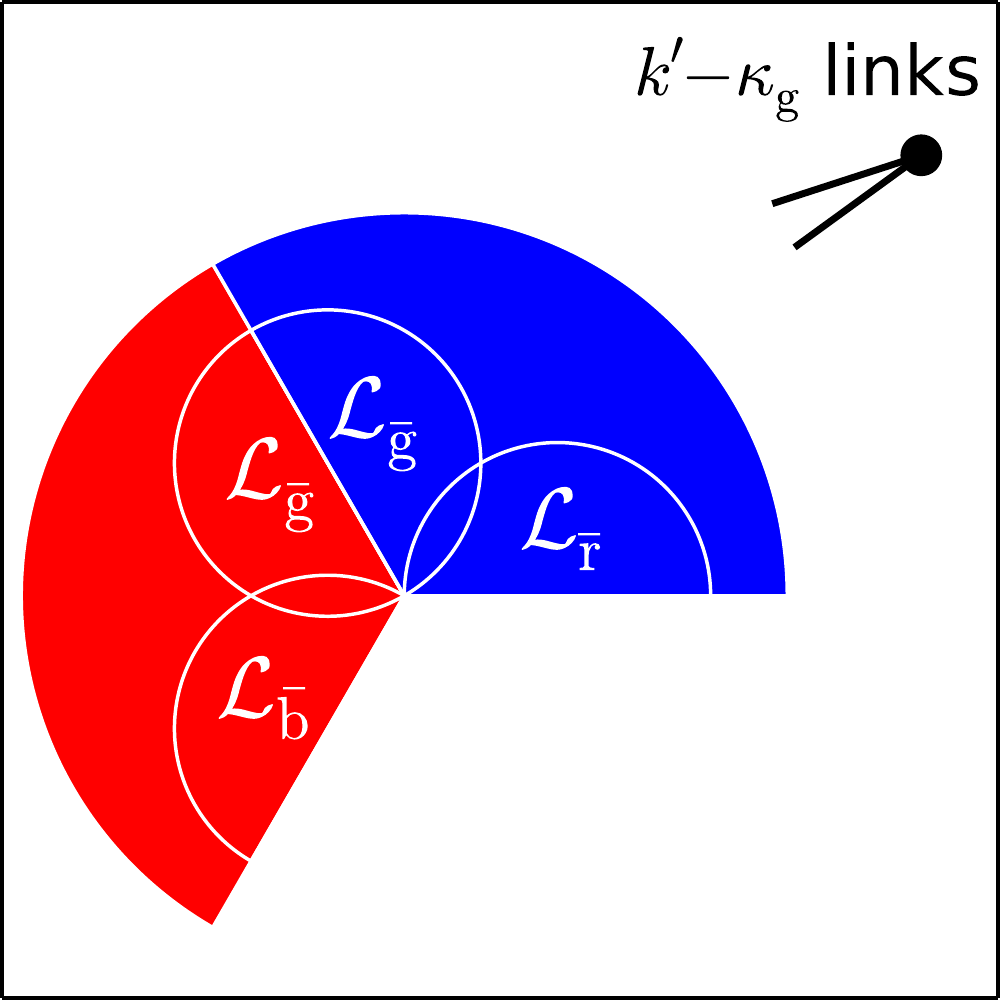}
     \hspace{-1mm}
    \includegraphics[trim=100 0 0 0,clip,height=0.4\columnwidth]{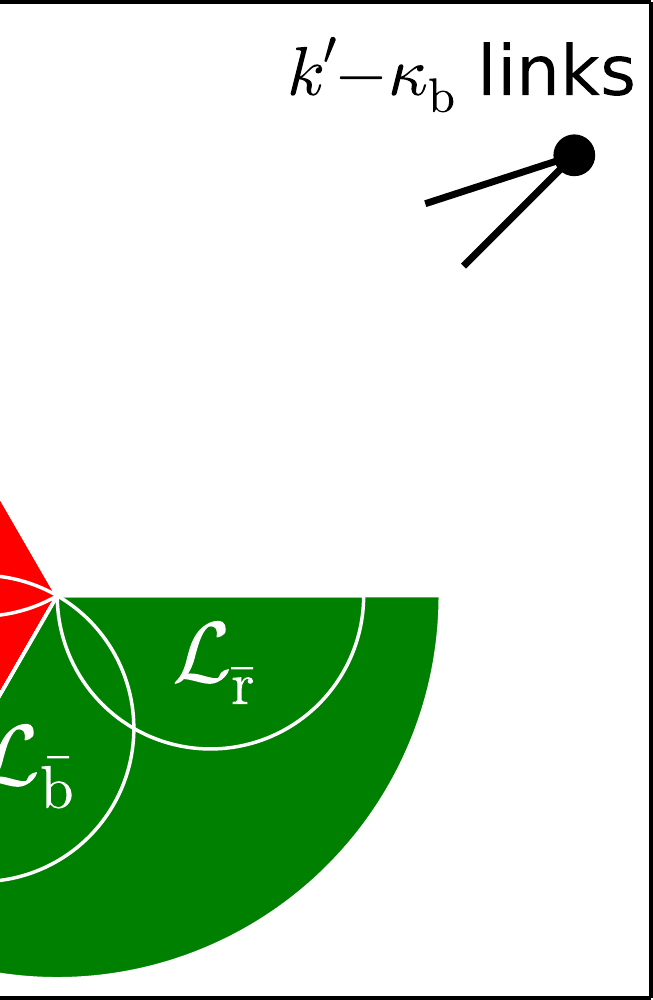}
     \hspace{-1mm}
    \includegraphics[trim=100 0 0 0,clip,height=0.4\columnwidth]{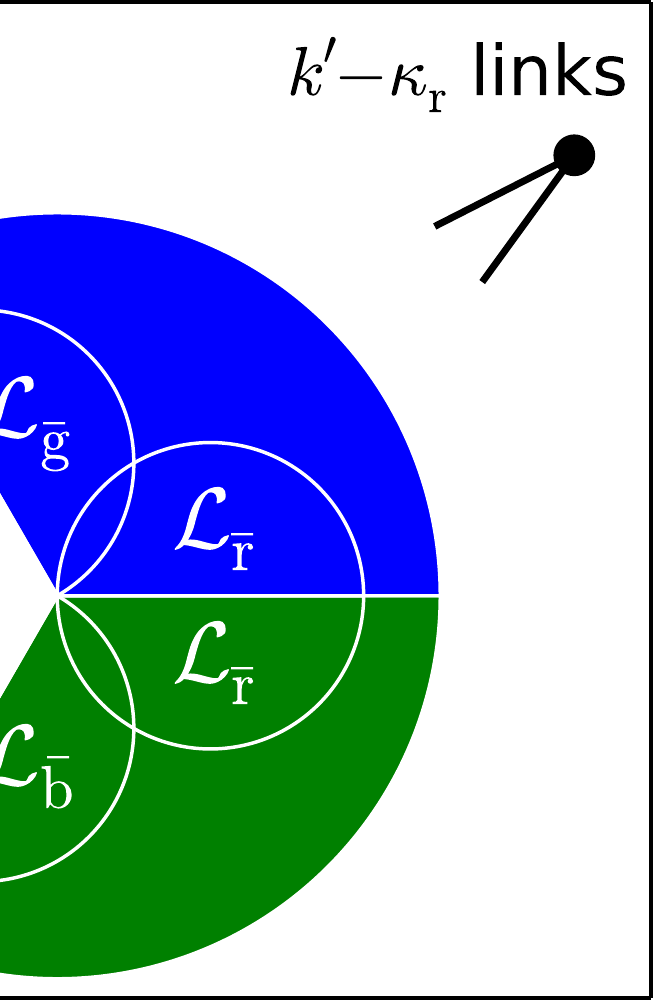}
   \end{center}
  \end{minipage}
    \caption{For calculating the probability of a node with $k$ links to belong to 
    $\mathcal{L}_{\rm color}$, we have to average over different link constellations which this node 
    might show. First, $B_{k,k'}$ is the probability that out of the $k$ links $k'$ 
    connect to the giant component. It is calculated using $u$ (compare figure~\ref{fig:primitives} 
    on the left). Second, $M_{k',\vec{\kappa}}$ gives the probability for a certain color 
    distribution among the links. It is calculated using $r_{\rm g}$ etc. (compare 
    figure~\ref{fig:primitives}, second from left). We assume that this second step is 
    independent of the first step, what is confirmed with the final results. Third, 
    $P_{\vec \kappa}$ gives the joint probability that for this color distribution 
    $\mathcal{L}_{\bar{\rm r}}$, $\mathcal{L}_{\bar{\rm b}}$ and $\mathcal{L}_{\bar{\rm g}}$ are connected to at 
    the same time. This is calculated using $U_{\bar{\rm r}}$ etc. (compare 
    figure~\ref{fig:primitives} on the right).}
    \label{fig:stepwise}
\end{figure}
As in equation~\ref{eq:S} for the giant component, 
we want to get an analytical result for $S_{\rm color}$ 
by averaging over possible link constellations of a randomly chosen node. 
Let us give the whole result and then explain it step by step afterwards:
\begin{align}
S_{\rm color} &= \sum_{k=0}^{\infty}p_k \sum_{k'=0}^{k} B_{k,k'} 
\sum_{\kappa_1,\dots, \kappa_C=0}^{k'} M_{k',\vec \kappa} 
P_{\vec \kappa},\label{eq:S_color}\tagS\\
B_{k,k'} &={k \choose k'}(1-u)^{k'}u^{k-k'},\label{eq:B}\tagS\\
M_{k',\vec \kappa} &=\frac{k'!}{\kappa_1! \times \dots \times \kappa_C!} \,
(r_1)^{\kappa_1} \times \dots \times (r_C)^{\kappa_C}\,
\delta_{k',\kappa_1+\dots + \kappa_C},\tagS\\
P_{\vec \kappa} &= \prod_{c=1}^C [1-(U_{\bar c})^{k'- \kappa_c }].\label{eq:p_success}\tagS
\end{align}
The formulas include the single link probabilities $r_c$, $u$ (equation~\eqref{eq:u}) and $U_{\bar C}$ 
(equation~\eqref{eq:U_c}~with~\eqref{eq:u_c}). 
An illustration of the procedure can be seen in figure~\ref{fig:stepwise}.
$B_{k,k'}$ is the binomial probability that out of the $k$ links $k'$ links connect to the giant component. 
$M_{k',\vec{\kappa}}$ gives the multinomial probability 
for a certain color distribution among the $k'$ links connecting to the giant component.
We assume that this second step is independent of the first step, 
what is confirmed with the final results. 
The numbers $\kappa_c$ count the links which connect to a node of color $c$ in the giant component. 
Finally, $P_{\vec \kappa}$ gives the joint probability 
that for the color distribution given by $\vec \kappa$ 
all giant $\mathcal{L}_{\bar c}$ are connected to at the same time. 
There is at least one link connecting to ${\mathcal L}_{\bar c}$ 
with probability $1-(U_{\bar c})^{k' - \kappa_c }$. 
The success probabilities for different colors have to be multiplied, 
as all ${\mathcal L}_{\bar c}$ have to be reached at the same time. 
We tested numerically that e.g. $U_{\bar{\rm 1}}$ and $U_{\bar{\rm 2}}$ are independent for a link 
connecting to the giant component and a third color.

\section{Examination of $S_{\rm color}$}

\subsection{Closed form solutions}

We now calculate closed form solutions for $S_{\rm color}$ for special cases. 
This is done to demonstrate how the extensive summations 
over $k'$, $k$ and $\vec \kappa$ can be performed analytically. 
In cases where this is not possible, 
a sampling of values $\vec \kappa$ has to be performed. 
The results can be tested against the analytically tractable situations 
and by comparing with numerical results. 
The closed form solutions presented here were used to calculate analytical 
results for the main article as well. 

For evaluating equation~\ref{eq:S_color} with two colors, 
we first rewrite
\begin{align}
\sigma_{k'} &\equiv \sum_{\kappa_1, \kappa_2=0}^{k'} M_{k',\vec \kappa} P_{\vec \kappa}\tagS\\
&= \sum_{\kappa_1=0}^{k'} {k' \choose \kappa_1}\, 
(r_1)^{\kappa_1} (r_2)^{k'-\kappa_1}\,
[1-(U_{\bar 1})^{k'-\kappa_1 }] [1-(U_{\bar 2})^{\kappa_1 }]\tagS\\
&= \sum_{\kappa_1=0}^{k'} {k' \choose \kappa_1}\, 
\left[
(r_1)^{\kappa_1} (r_2)^{k'-\kappa_1}\, - 
(r_1 U_{\bar 2})^{\kappa_1} (r_2)^{k'-\kappa_1}\,-
(r_1)^{\kappa_1} (r_2 U_{\bar 1})^{k'-\kappa_1}\,+
(r_1 U_{\bar 2})^{\kappa_1} (r_2 U_{\bar 1})^{k'-\kappa_1}\right]\tagS\\
 &= 1 - (r_1+r_2 U_{\bar 1})^{k'} - (r_2+r_1 U_{\bar 2})^{k'} + (r_1 U_{\bar 2} + r_2 U_{\bar 1})^{k'}.\tagS
\end{align}
In the last step, the binomial formula was used backward. 
We can use this procedure once more, and with equation~\ref{eq:U_c} and $r_1+r_2=1$ we find 
\begin{align}
S_{\rm color} &= \sum_k p_k \sum_{k'=0}^{k} B_{k,k'} \sigma_{k'} \tagS\\
 &= \sum_k p_k \sum_{k'=0}^{k} {k \choose k'}u^{k-k'}\, 
\left[
(1-u)^{k'} - ((1-u)(r_1+r_2U_{\bar 1}))^{k'} - \dots
\right]\tagS\\
 &= \sum_k p_k \left[1 - (u_{\bar 1})^k - (u_{\bar 2})^k + (u_{\bar 1}+u_{\bar 2}-1)^k\right]\tagS\\
 &= 1 - g_0(u_{\bar 1}) - g_0(u_{\bar 2}) + g_0(u_{\bar 1}+u_{\bar 2}-1).\label{eq:two_colors}\tagS
\end{align}
This result holds for any degree distribution and color distribution. 
Notice that $r_c\leq u_{\bar c}\leq 1$. 
The result for two colors does only depend on the probabilities $u_{\bar c}$, 
while conditional probabilities as $U_{\bar c}$ were eliminated. 
This was possible as $\mathcal{L}_{\bar 1}$ and $\mathcal{L}_{\bar 2}$ 
are not overlapping for two colors. 
For Poisson graphs we find with the according generating function 
\begin{align}
g_0(z) &= g_1(z) = e^{\bar{k}(z-1)}.\label{eq:g0}\tagS\\
S_{\rm color} &= [1 - g_0(u_{\bar 1})][1 - g_0(u_{\bar 2})].\tagS
\end{align}
For more than two colors, $\mathcal{L}_{\bar c}$ do overlap. 
For homogeneous color distributions $r_c=1/C$, a closed form solution can be 
found in the same way as for two colors with the binomial formula. We find 
\begin{align}
S_{\rm color} &=  \sum_{j=0}^C (-1)^j {C \choose j} g_0\left[u+(1-u)\left(\frac{j}{C}U_{\bar 1}^{j-1} + \frac{C-j}{C}U_{\bar 1}^{j}\right)\right].\tagS
\end{align}

Let us finally discuss the behavior for $C\to \infty$. This can be done utilizing 
the term $\sigma_{k'}$, the probability that a node connecting over $k'$ links to the giant component 
belongs to $S_{\rm color}$. 
As can be seen with equation~\ref{eq:p_success}, $\sigma_0=\sigma_1=0$. On the other hand, 
with $r_c\to 0$, equation~\ref{eq:u_c} converges to equation~\ref{eq:u} and therefore $U_{\bar 1}\to 0$. 
This means that $\sigma_{k'>1}\to 1$. We finally find with equation~\ref{eq:S_color}
\begin{align}
S_{\rm color,\infty} &\equiv \lim_{C\to \infty} S_{\rm color}\tagS\\
 &= 1-\sum_{k=0}^{\infty}p_k [u^k +k (1-u) u^{k-1}]\tagS\\
 &= 1 - g_0(u) - (1-u) \left.\frac{{\rm d}g_0(z)}{{\rm d}z}\right|_{z=u}.\tagS
\end{align}

\subsection{Critical behavior for Poisson graphs}

With equation~\ref{eq:S_color}, vanishing $\sigma_{k'}$
causes $S_{\rm color}=0$. According to 
\begin{align}
\sigma_{k'} &= \sum_{\kappa_1,\dots, \kappa_C=0}^{k'} M_{k',\vec \kappa} 
\prod_{c=1}^C [1-(U_{\bar c})^{k'- \kappa_c }].\label{eq:sigmak}\tagS
\end{align}
this is the case if $U_{\bar c}=1$ for any color $c$. 
With equation~\ref{eq:U_c} we find that $U_{\bar c}=1$ whenever $u_{\bar c}=1$. 
Examining equation~\ref{eq:u_c} for $u_{\bar c}$, 
we can relate to site percolation (random removal of nodes). 
For Poisson graphs we have $r_{\rm crit}=(\bar{k}-1)/\bar{k}$.
With homogeneous color distribution $r_c=1/C$, we can resolve the critical connectivity 
given the number of colors
\begin{align}
\bar{k}_{\rm crit} &= C/(C-1). \tagS
\end{align}

The normal giant component size $S$ shows a special critical behavior shortly above the transition point, 
it scales linearly with $\bar{k}-1$. Here we are interested in the behavior of $S_{\rm color}$ 
which is a function of $1-u_{\bar 1}$ which itself can be related to $1-u=S$. 
With inserting into equation~\ref{eq:u_c} it can be shown that 
$u_{\bar c}({\bar k})=r_c+(1-r_c)u((1-r_c){\bar k})$. For small arguments
$(\bar{k}-{\bar k}_{\rm crit})$, 
\begin{align}
1-u_{\bar 1}(\bar{k}>{\bar k}_{\rm crit}) &\approx 
(1-r_1)^2 \left.\frac{{\rm d}(1-u)}{{\rm d}{\bar k}}\right|_{\bar{k}=1+0} (\bar{k}-{\bar k}_{\rm crit}).\tagS
\end{align}
Inserting into equation~\ref{eq:U_c} we find using 
$1-u(\bar{k}>1) \approx \left.\frac{{\rm d}(1-u)}{{\rm d}{\bar k}}\right|_{\bar{k}=1+0} (\bar{k}-1)$
\begin{align}
\varepsilon &\equiv 1 - U_{\bar 1} \approx C (\bar{k}-{\bar k}_{\rm crit})\tagS
\end{align}
if additionally $\bar{k}-{\bar k}_{\rm crit}\ll\bar{k}-1$ holds ($1-u_{\bar 1}$ small compared to 
$1-u$). 

For calculating $\sigma_{k'}$, we first need to evaluate $P_{\vec \kappa}$ including expressions 
$1-(U_{\bar 1})^{k'-\kappa_c}$. Replacing with $\varepsilon$ and applying an approximation we find 
$1-(U_{\bar 1})^{k'-\kappa_c}=1-(1-\varepsilon)^{k'-\kappa_c}\approx (k'-\kappa_c) \varepsilon$. 
This is true at least as long as $k'\varepsilon\ll 1$. With this we find 
$P_{\vec \kappa}\propto (\bar{k}-{\bar k}_{\rm crit})^C$ independent of $\vec \kappa$, and finally 
\begin{align}
\sigma_{k'} &\propto (\bar{k}-{\bar k}_{\rm crit})^C. \tagS
\end{align}
We finally find 
\begin{align}
S_{\rm color} &\propto (\bar{k}-{\bar k}_{\rm crit})^{\beta},\tagS\\
\beta &= C.\tagS
\end{align}

The critical behavior of $S_{\rm color,\infty}$ for Poisson graphs can be evaluated with the generating function eq.~(\ref{eq:g0}) and $S=1-u$. 
We find $S_{\rm color,\infty}=S-{\bar k} S (1-S)$, and for small positive ${\bar k}-1$ the giant component grows approximately with $S\approx 2 ({\bar k}-1)/{\bar k}^2$. Therefore
\begin{align}
S_{\rm color,\infty}\approx ({\bar k}-1)^2(4/{\bar k}^2-2/{\bar k}^3)\propto ({\bar k}-1)^2.\tagS
\end{align}

\bibliography{avoidable_colors_paper}

\begin{thebibliography}{42}
\expandafter\ifx\csname natexlab\endcsname\relax\def\natexlab#1{#1}\fi
\expandafter\ifx\csname bibnamefont\endcsname\relax
  \def\bibnamefont#1{#1}\fi
\expandafter\ifx\csname bibfnamefont\endcsname\relax
  \def\bibfnamefont#1{#1}\fi
\expandafter\ifx\csname citenamefont\endcsname\relax
  \def\citenamefont#1{#1}\fi
\expandafter\ifx\csname url\endcsname\relax
  \def\url#1{\texttt{#1}}\fi
\expandafter\ifx\csname urlprefix\endcsname\relax\def\urlprefix{URL }\fi
\providecommand{\bibinfo}[2]{#2}
\providecommand{\eprint}[2][]{\url{#2}}

\bibitem[{\citenamefont{Stelling et~al.}(2004)\citenamefont{Stelling, Sauer,
  Szallasi, III, and Doyle}}]{stelling-cell2004}
\bibinfo{author}{\bibfnamefont{J.}~\bibnamefont{Stelling}},
  \bibinfo{author}{\bibfnamefont{U.}~\bibnamefont{Sauer}},
  \bibinfo{author}{\bibfnamefont{Z.}~\bibnamefont{Szallasi}},
  \bibinfo{author}{\bibfnamefont{F.~J.~D.} \bibnamefont{III}},
  \bibnamefont{and} \bibinfo{author}{\bibfnamefont{J.}~\bibnamefont{Doyle}},
  \bibinfo{journal}{Cell} \textbf{\bibinfo{volume}{118}}, \bibinfo{pages}{675 }
  (\bibinfo{year}{2004}), ISSN \bibinfo{issn}{0092-8674},
  \urlprefix\url{http://www.sciencedirect.com/science/article/pii/S00928674040%
08402}.

\bibitem[{\citenamefont{Carmi et~al.}(2007)\citenamefont{Carmi, Havlin,
  Kirkpatrick, Shavitt, and Shir}}]{carmi-pnas2007}
\bibinfo{author}{\bibfnamefont{S.}~\bibnamefont{Carmi}},
  \bibinfo{author}{\bibfnamefont{S.}~\bibnamefont{Havlin}},
  \bibinfo{author}{\bibfnamefont{S.}~\bibnamefont{Kirkpatrick}},
  \bibinfo{author}{\bibfnamefont{Y.}~\bibnamefont{Shavitt}}, \bibnamefont{and}
  \bibinfo{author}{\bibfnamefont{E.}~\bibnamefont{Shir}},
  \bibinfo{journal}{Proceedings of the National Academy of Sciences}
  \textbf{\bibinfo{volume}{104}}, \bibinfo{pages}{11150}
  (\bibinfo{year}{2007}),
  \urlprefix\url{http://www.pnas.org/content/104/27/11150.abstract}.

\bibitem[{\citenamefont{Neumayer et~al.}(2008)\citenamefont{Neumayer, Zussman,
  Cohen, and Modiano}}]{neumayer-milcom2008}
\bibinfo{author}{\bibfnamefont{S.}~\bibnamefont{Neumayer}},
  \bibinfo{author}{\bibfnamefont{G.}~\bibnamefont{Zussman}},
  \bibinfo{author}{\bibfnamefont{R.}~\bibnamefont{Cohen}}, \bibnamefont{and}
  \bibinfo{author}{\bibfnamefont{E.}~\bibnamefont{Modiano}}, in
  \emph{\bibinfo{booktitle}{{Military Communications Conference, 2008. MILCOM
  2008. IEEE}}} (\bibinfo{year}{2008}), pp. \bibinfo{pages}{1--6}.

\bibitem[{\citenamefont{Agarwal et~al.}(2011)\citenamefont{Agarwal, Efrat,
  Ganjugunte, Hay, Sankararaman, and Zussman}}]{agarwal-infocom2011}
\bibinfo{author}{\bibfnamefont{P.~K.} \bibnamefont{Agarwal}},
  \bibinfo{author}{\bibfnamefont{A.}~\bibnamefont{Efrat}},
  \bibinfo{author}{\bibfnamefont{S.}~\bibnamefont{Ganjugunte}},
  \bibinfo{author}{\bibfnamefont{D.}~\bibnamefont{Hay}},
  \bibinfo{author}{\bibfnamefont{S.}~\bibnamefont{Sankararaman}},
  \bibnamefont{and} \bibinfo{author}{\bibfnamefont{G.}~\bibnamefont{Zussman}},
  in \emph{\bibinfo{booktitle}{{INFOCOM, 2011 Proceedings IEEE}}}
  (\bibinfo{year}{2011}), pp. \bibinfo{pages}{1521--1529}, ISSN
  \bibinfo{issn}{0743-166X}.

\bibitem[{\citenamefont{{Berezin} et~al.}(2015)\citenamefont{{Berezin},
  {Bashan}, {Danziger}, {Li}, and {Havlin}}}]{berezin-scireps2015}
\bibinfo{author}{\bibfnamefont{Y.}~\bibnamefont{{Berezin}}},
  \bibinfo{author}{\bibfnamefont{A.}~\bibnamefont{{Bashan}}},
  \bibinfo{author}{\bibfnamefont{M.~M.} \bibnamefont{{Danziger}}},
  \bibinfo{author}{\bibfnamefont{D.}~\bibnamefont{{Li}}}, \bibnamefont{and}
  \bibinfo{author}{\bibfnamefont{S.}~\bibnamefont{{Havlin}}},
  \bibinfo{journal}{Scientific Reports} \textbf{\bibinfo{volume}{5}},
  \bibinfo{pages}{8934} (\bibinfo{year}{2015}),
  \urlprefix\url{http://www.nature.com/srep/2015/150311/srep08934/full/srep089%
34.html}.

\bibitem[{\citenamefont{{Schuster Stefan} et~al.}(2000)\citenamefont{{Schuster
  Stefan}, {Fell David A.}, and {Dandekar Thomas}}}]{Schuster2000Metabolic}
\bibinfo{author}{\bibnamefont{{Schuster Stefan}}},
  \bibinfo{author}{\bibnamefont{{Fell David A.}}}, \bibnamefont{and}
  \bibinfo{author}{\bibnamefont{{Dandekar Thomas}}}, \bibinfo{journal}{Nat
  Biotech} \textbf{\bibinfo{volume}{18}}, \bibinfo{pages}{326}
  (\bibinfo{year}{2000}), ISSN \bibinfo{issn}{1087-0156},
  \bibinfo{note}{10.1038/73786}.

\bibitem[{\citenamefont{{Feil Robert} and {Fraga Mario
  F.}}(2012)}]{Feil2012epigenetics}
\bibinfo{author}{\bibnamefont{{Feil Robert}}} \bibnamefont{and}
  \bibinfo{author}{\bibnamefont{{Fraga Mario F.}}}, \bibinfo{journal}{Nat Rev
  Genet} \textbf{\bibinfo{volume}{13}}, \bibinfo{pages}{97}
  (\bibinfo{year}{2012}), ISSN \bibinfo{issn}{1471-0056},
  \bibinfo{note}{10.1038/nrg3142}.

\bibitem[{\citenamefont{P\'{a}l et~al.}(2006)\citenamefont{P\'{a}l, Papp,
  Lercher, Csermely, Oliver, and Hurst}}]{pal-nature2006}
\bibinfo{author}{\bibfnamefont{C.}~\bibnamefont{P\'{a}l}},
  \bibinfo{author}{\bibfnamefont{B.}~\bibnamefont{Papp}},
  \bibinfo{author}{\bibfnamefont{M.~J.} \bibnamefont{Lercher}},
  \bibinfo{author}{\bibfnamefont{P.}~\bibnamefont{Csermely}},
  \bibinfo{author}{\bibfnamefont{S.~G.} \bibnamefont{Oliver}},
  \bibnamefont{and} \bibinfo{author}{\bibfnamefont{L.~D.} \bibnamefont{Hurst}},
  \bibinfo{journal}{Nature} \textbf{\bibinfo{volume}{440}},
  \bibinfo{pages}{667–670} (\bibinfo{year}{2006}), ISSN
  \bibinfo{issn}{1476-4679},
  \urlprefix\url{http://dx.doi.org/10.1038/nature04568}.

\bibitem[{\citenamefont{White et~al.}(2013)\citenamefont{White, Gerdin, Karp,
  Ryder, Buljan, Bussell, Salisbury, Clare, Ingham, Podrini
  et~al.}}]{white-cell2013}
\bibinfo{author}{\bibfnamefont{J.~K.} \bibnamefont{White}},
  \bibinfo{author}{\bibfnamefont{A.-K.} \bibnamefont{Gerdin}},
  \bibinfo{author}{\bibfnamefont{N.~A.} \bibnamefont{Karp}},
  \bibinfo{author}{\bibfnamefont{E.}~\bibnamefont{Ryder}},
  \bibinfo{author}{\bibfnamefont{M.}~\bibnamefont{Buljan}},
  \bibinfo{author}{\bibfnamefont{J.}~\bibnamefont{Bussell}},
  \bibinfo{author}{\bibfnamefont{J.}~\bibnamefont{Salisbury}},
  \bibinfo{author}{\bibfnamefont{S.}~\bibnamefont{Clare}},
  \bibinfo{author}{\bibfnamefont{N.~J.} \bibnamefont{Ingham}},
  \bibinfo{author}{\bibfnamefont{C.}~\bibnamefont{Podrini}},
  \bibnamefont{et~al.}, \bibinfo{journal}{Cell} \textbf{\bibinfo{volume}{154}},
  \bibinfo{pages}{452 } (\bibinfo{year}{2013}), ISSN \bibinfo{issn}{0092-8674},
  \urlprefix\url{http://www.sciencedirect.com/science/article/pii/S00928674130%
07617}.

\bibitem[{\citenamefont{Zallen}(1977)}]{zallen-prb1977}
\bibinfo{author}{\bibfnamefont{R.}~\bibnamefont{Zallen}},
  \bibinfo{journal}{Phys. Rev. B} \textbf{\bibinfo{volume}{16}},
  \bibinfo{pages}{1426} (\bibinfo{year}{1977}),
  \urlprefix\url{http://link.aps.org/doi/10.1103/PhysRevB.16.1426}.

\bibitem[{\citenamefont{Wierman}(1989)}]{wierman-banach1989}
\bibinfo{author}{\bibfnamefont{J.~C.} \bibnamefont{Wierman}},
  \bibinfo{journal}{Banach Center Publications} \textbf{\bibinfo{volume}{25}},
  \bibinfo{pages}{241} (\bibinfo{year}{1989}).

\bibitem[{\citenamefont{Peixoto}(2014)}]{peixoto2014hierarchical}
\bibinfo{author}{\bibfnamefont{T.~P.} \bibnamefont{Peixoto}},
  \bibinfo{journal}{Physical Review X} \textbf{\bibinfo{volume}{4}},
  \bibinfo{pages}{011047} (\bibinfo{year}{2014}).

\bibitem[{Cai()}]{CaidaData}
\emph{\bibinfo{title}{Caida project}}, \bibinfo{howpublished}{\url
  {http://www.caida.org/data/}}.

\bibitem[{\citenamefont{Cohen and Havlin}(2010)}]{cohen-book2010}
\bibinfo{author}{\bibfnamefont{R.}~\bibnamefont{Cohen}} \bibnamefont{and}
  \bibinfo{author}{\bibfnamefont{S.}~\bibnamefont{Havlin}},
  \emph{\bibinfo{title}{Complex Networks: Structure, Robustness and Function}}
  (\bibinfo{publisher}{Cambridge University Press}, \bibinfo{year}{2010}), ISBN
  \bibinfo{isbn}{9781139489270}.

\bibitem[{\citenamefont{Newman}(2010)}]{newman-book2010}
\bibinfo{author}{\bibfnamefont{M.}~\bibnamefont{Newman}},
  \emph{\bibinfo{title}{{Networks: an introduction}}} (\bibinfo{publisher}{OUP
  Oxford}, \bibinfo{year}{2010}).

\bibitem[{\citenamefont{Boccaletti et~al.}(2006)\citenamefont{Boccaletti,
  Latora, Moreno, Chavez, and Hwang}}]{boccaletti-physicsreports2006}
\bibinfo{author}{\bibfnamefont{S.}~\bibnamefont{Boccaletti}},
  \bibinfo{author}{\bibfnamefont{V.}~\bibnamefont{Latora}},
  \bibinfo{author}{\bibfnamefont{Y.}~\bibnamefont{Moreno}},
  \bibinfo{author}{\bibfnamefont{M.}~\bibnamefont{Chavez}}, \bibnamefont{and}
  \bibinfo{author}{\bibfnamefont{D.-U.} \bibnamefont{Hwang}},
  \bibinfo{journal}{Physics Reports} \textbf{\bibinfo{volume}{424}},
  \bibinfo{pages}{175 } (\bibinfo{year}{2006}), ISSN \bibinfo{issn}{0370-1573},
  \urlprefix\url{http://www.sciencedirect.com/science/article/pii/S03701573050%
0462X}.

\bibitem[{\citenamefont{Caldarelli}(2007)}]{caldarelli-sfbook2007}
\bibinfo{author}{\bibfnamefont{G.}~\bibnamefont{Caldarelli}},
  \emph{\bibinfo{title}{Scale-free networks: complex webs in nature and
  technology}} (\bibinfo{publisher}{Oxford University Press},
  \bibinfo{year}{2007}).

\bibitem[{\citenamefont{Achlioptas et~al.}(2009)\citenamefont{Achlioptas,
  D'Souza, and Spencer}}]{achlioptas-science2009}
\bibinfo{author}{\bibfnamefont{D.}~\bibnamefont{Achlioptas}},
  \bibinfo{author}{\bibfnamefont{R.~M.} \bibnamefont{D'Souza}},
  \bibnamefont{and} \bibinfo{author}{\bibfnamefont{J.}~\bibnamefont{Spencer}},
  \bibinfo{journal}{Science} \textbf{\bibinfo{volume}{323}},
  \bibinfo{pages}{1453} (\bibinfo{year}{2009}),
  \urlprefix\url{http://www.sciencemag.org/content/323/5920/1453.abstract}.

\bibitem[{\citenamefont{Cohen et~al.}(2000)\citenamefont{Cohen, Erez,
  Ben-Avraham, and Havlin}}]{cohen-2000resilience}
\bibinfo{author}{\bibfnamefont{R.}~\bibnamefont{Cohen}},
  \bibinfo{author}{\bibfnamefont{K.}~\bibnamefont{Erez}},
  \bibinfo{author}{\bibfnamefont{D.}~\bibnamefont{Ben-Avraham}},
  \bibnamefont{and} \bibinfo{author}{\bibfnamefont{S.}~\bibnamefont{Havlin}},
  \bibinfo{journal}{Physical review letters} \textbf{\bibinfo{volume}{85}},
  \bibinfo{pages}{4626} (\bibinfo{year}{2000}).

\bibitem[{\citenamefont{Cohen et~al.}(2001)\citenamefont{Cohen, Erez, ben
  Avraham, and Havlin}}]{cohen-prl2001}
\bibinfo{author}{\bibfnamefont{R.}~\bibnamefont{Cohen}},
  \bibinfo{author}{\bibfnamefont{K.}~\bibnamefont{Erez}},
  \bibinfo{author}{\bibfnamefont{D.}~\bibnamefont{ben Avraham}},
  \bibnamefont{and} \bibinfo{author}{\bibfnamefont{S.}~\bibnamefont{Havlin}},
  \bibinfo{journal}{Phys. Rev. Lett.} \textbf{\bibinfo{volume}{86}},
  \bibinfo{pages}{3682} (\bibinfo{year}{2001}),
  \urlprefix\url{http://link.aps.org/doi/10.1103/PhysRevLett.86.3682}.

\bibitem[{\citenamefont{Pastor-Satorras and
  Vespignani}(2001)}]{pastorsatorras-prl2001}
\bibinfo{author}{\bibfnamefont{R.}~\bibnamefont{Pastor-Satorras}}
  \bibnamefont{and}
  \bibinfo{author}{\bibfnamefont{A.}~\bibnamefont{Vespignani}},
  \bibinfo{journal}{Phys. Rev. Lett.} \textbf{\bibinfo{volume}{86}},
  \bibinfo{pages}{3200} (\bibinfo{year}{2001}),
  \urlprefix\url{http://link.aps.org/doi/10.1103/PhysRevLett.86.3200}.

\bibitem[{\citenamefont{Sasson et~al.}(2003)\citenamefont{Sasson, Cavin, and
  Schiper}}]{sasson2003probabilistic}
\bibinfo{author}{\bibfnamefont{Y.}~\bibnamefont{Sasson}},
  \bibinfo{author}{\bibfnamefont{D.}~\bibnamefont{Cavin}}, \bibnamefont{and}
  \bibinfo{author}{\bibfnamefont{A.}~\bibnamefont{Schiper}}, in
  \emph{\bibinfo{booktitle}{Wireless Communications and Networking, 2003. WCNC
  2003. 2003 IEEE}} (\bibinfo{organization}{IEEE}, \bibinfo{year}{2003}),
  vol.~\bibinfo{volume}{2}, pp. \bibinfo{pages}{1124--1130}.

\bibitem[{\citenamefont{Goldenberg et~al.}(2000)\citenamefont{Goldenberg,
  Libai, Solomon, Jan, and Stauffer}}]{goldenberg-physa2000}
\bibinfo{author}{\bibfnamefont{J.}~\bibnamefont{Goldenberg}},
  \bibinfo{author}{\bibfnamefont{B.}~\bibnamefont{Libai}},
  \bibinfo{author}{\bibfnamefont{S.}~\bibnamefont{Solomon}},
  \bibinfo{author}{\bibfnamefont{N.}~\bibnamefont{Jan}}, \bibnamefont{and}
  \bibinfo{author}{\bibfnamefont{D.}~\bibnamefont{Stauffer}},
  \bibinfo{journal}{Physica A: Statistical Mechanics and its Applications}
  \textbf{\bibinfo{volume}{284}}, \bibinfo{pages}{335 } (\bibinfo{year}{2000}),
  ISSN \bibinfo{issn}{0378-4371},
  \urlprefix\url{http://www.sciencedirect.com/science/article/pii/S03784371000%
02600}.

\bibitem[{\citenamefont{Solomon et~al.}(2000)\citenamefont{Solomon, Weisbuch,
  de~Arcangelis, Jan, and Stauffer}}]{solomon-physa2000}
\bibinfo{author}{\bibfnamefont{S.}~\bibnamefont{Solomon}},
  \bibinfo{author}{\bibfnamefont{G.}~\bibnamefont{Weisbuch}},
  \bibinfo{author}{\bibfnamefont{L.}~\bibnamefont{de~Arcangelis}},
  \bibinfo{author}{\bibfnamefont{N.}~\bibnamefont{Jan}}, \bibnamefont{and}
  \bibinfo{author}{\bibfnamefont{D.}~\bibnamefont{Stauffer}},
  \bibinfo{journal}{Physica A: Statistical Mechanics and its Applications}
  \textbf{\bibinfo{volume}{277}}, \bibinfo{pages}{239 } (\bibinfo{year}{2000}),
  ISSN \bibinfo{issn}{0378-4371},
  \urlprefix\url{http://www.sciencedirect.com/science/article/pii/S03784371990%
05439}.

\bibitem[{\citenamefont{Breskin et~al.}(2006)\citenamefont{Breskin, Soriano,
  Moses, and Tlusty}}]{breskin-prl2006}
\bibinfo{author}{\bibfnamefont{I.}~\bibnamefont{Breskin}},
  \bibinfo{author}{\bibfnamefont{J.}~\bibnamefont{Soriano}},
  \bibinfo{author}{\bibfnamefont{E.}~\bibnamefont{Moses}}, \bibnamefont{and}
  \bibinfo{author}{\bibfnamefont{T.}~\bibnamefont{Tlusty}},
  \bibinfo{journal}{Phys. Rev. Lett.} \textbf{\bibinfo{volume}{97}},
  \bibinfo{pages}{188102} (\bibinfo{year}{2006}),
  \urlprefix\url{http://link.aps.org/doi/10.1103/PhysRevLett.97.188102}.

\bibitem[{\citenamefont{Smart et~al.}(2008)\citenamefont{Smart, Amaral, and
  Ottino}}]{smart-pnas2008}
\bibinfo{author}{\bibfnamefont{A.~G.} \bibnamefont{Smart}},
  \bibinfo{author}{\bibfnamefont{L.~A.~N.} \bibnamefont{Amaral}},
  \bibnamefont{and} \bibinfo{author}{\bibfnamefont{J.~M.}
  \bibnamefont{Ottino}}, \bibinfo{journal}{Proceedings of the National Academy
  of Sciences} \textbf{\bibinfo{volume}{105}}, \bibinfo{pages}{13223}
  (\bibinfo{year}{2008}),
  \eprint{http://www.pnas.org/content/105/36/13223.full.pdf+html},
  \urlprefix\url{http://www.pnas.org/content/105/36/13223.abstract}.

\bibitem[{\citenamefont{Aon et~al.}(2004)\citenamefont{Aon, Cortassa, and
  O'Rourke}}]{aon-pnas2004}
\bibinfo{author}{\bibfnamefont{M.~A.} \bibnamefont{Aon}},
  \bibinfo{author}{\bibfnamefont{S.}~\bibnamefont{Cortassa}}, \bibnamefont{and}
  \bibinfo{author}{\bibfnamefont{B.}~\bibnamefont{O'Rourke}},
  \bibinfo{journal}{Proceedings of the National Academy of Sciences of the
  United States of America} \textbf{\bibinfo{volume}{101}},
  \bibinfo{pages}{4447} (\bibinfo{year}{2004}),
  \eprint{http://www.pnas.org/content/101/13/4447.full.pdf+html},
  \urlprefix\url{http://www.pnas.org/content/101/13/4447.abstract}.

\bibitem[{\citenamefont{Dorogovtsev et~al.}(2006)\citenamefont{Dorogovtsev,
  Goltsev, and Mendes}}]{dorogovtsev-prl2006}
\bibinfo{author}{\bibfnamefont{S.~N.} \bibnamefont{Dorogovtsev}},
  \bibinfo{author}{\bibfnamefont{A.~V.} \bibnamefont{Goltsev}},
  \bibnamefont{and} \bibinfo{author}{\bibfnamefont{J.~F.~F.}
  \bibnamefont{Mendes}}, \bibinfo{journal}{Phys. Rev. Lett.}
  \textbf{\bibinfo{volume}{96}}, \bibinfo{pages}{040601}
  (\bibinfo{year}{2006}),
  \urlprefix\url{http://link.aps.org/doi/10.1103/PhysRevLett.96.040601}.

\bibitem[{\citenamefont{Goltsev et~al.}(2006)\citenamefont{Goltsev,
  Dorogovtsev, and Mendes}}]{goltsev-pre2006}
\bibinfo{author}{\bibfnamefont{A.~V.} \bibnamefont{Goltsev}},
  \bibinfo{author}{\bibfnamefont{S.~N.} \bibnamefont{Dorogovtsev}},
  \bibnamefont{and} \bibinfo{author}{\bibfnamefont{J.~F.~F.}
  \bibnamefont{Mendes}}, \bibinfo{journal}{Phys. Rev. E}
  \textbf{\bibinfo{volume}{73}}, \bibinfo{pages}{056101}
  (\bibinfo{year}{2006}),
  \urlprefix\url{http://link.aps.org/doi/10.1103/PhysRevE.73.056101}.

\bibitem[{\citenamefont{Buldyrev et~al.}(2010)\citenamefont{Buldyrev, Parshani,
  Paul, Stanley, and Havlin}}]{buldyrev-nature2010}
\bibinfo{author}{\bibfnamefont{S.~V.} \bibnamefont{Buldyrev}},
  \bibinfo{author}{\bibfnamefont{R.}~\bibnamefont{Parshani}},
  \bibinfo{author}{\bibfnamefont{G.}~\bibnamefont{Paul}},
  \bibinfo{author}{\bibfnamefont{H.~E.} \bibnamefont{Stanley}},
  \bibnamefont{and} \bibinfo{author}{\bibfnamefont{S.}~\bibnamefont{Havlin}},
  \bibinfo{journal}{Nature} \textbf{\bibinfo{volume}{464}},
  \bibinfo{pages}{1025} (\bibinfo{year}{2010}),
  \urlprefix\url{http://dx.doi.org/10.1038/nature08932}.

\bibitem[{\citenamefont{Baxter et~al.}(2012)\citenamefont{Baxter, Dorogovtsev,
  Goltsev, and Mendes}}]{baxter-prl2012}
\bibinfo{author}{\bibfnamefont{G.~J.} \bibnamefont{Baxter}},
  \bibinfo{author}{\bibfnamefont{S.~N.} \bibnamefont{Dorogovtsev}},
  \bibinfo{author}{\bibfnamefont{A.~V.} \bibnamefont{Goltsev}},
  \bibnamefont{and} \bibinfo{author}{\bibfnamefont{J.~F.~F.}
  \bibnamefont{Mendes}}, \bibinfo{journal}{Phys. Rev. Lett.}
  \textbf{\bibinfo{volume}{109}}, \bibinfo{pages}{248701}
  (\bibinfo{year}{2012}),
  \urlprefix\url{http://link.aps.org/doi/10.1103/PhysRevLett.109.248701}.

\bibitem[{\citenamefont{Boccaletti et~al.}(2014)\citenamefont{Boccaletti,
  Bianconi, Criado, Del~Genio, G{\'o}mez-Garde{\~n}es, Romance, Sendina-Nadal,
  Wang, and Zanin}}]{boccaletti-physicsreports2014}
\bibinfo{author}{\bibfnamefont{S.}~\bibnamefont{Boccaletti}},
  \bibinfo{author}{\bibfnamefont{G.}~\bibnamefont{Bianconi}},
  \bibinfo{author}{\bibfnamefont{R.}~\bibnamefont{Criado}},
  \bibinfo{author}{\bibfnamefont{C.}~\bibnamefont{Del~Genio}},
  \bibinfo{author}{\bibfnamefont{J.}~\bibnamefont{G{\'o}mez-Garde{\~n}es}},
  \bibinfo{author}{\bibfnamefont{M.}~\bibnamefont{Romance}},
  \bibinfo{author}{\bibfnamefont{I.}~\bibnamefont{Sendina-Nadal}},
  \bibinfo{author}{\bibfnamefont{Z.}~\bibnamefont{Wang}}, \bibnamefont{and}
  \bibinfo{author}{\bibfnamefont{M.}~\bibnamefont{Zanin}},
  \bibinfo{journal}{Physics Reports}  (\bibinfo{year}{2014}), ISSN
  \bibinfo{issn}{0370-1573},
  \urlprefix\url{http://dx.doi.org/10.1016/j.physrep.2014.07.001}.

\bibitem[{\citenamefont{Erd\H{o}s and R\'{e}nyi}(1959)}]{erd-1959random}
\bibinfo{author}{\bibfnamefont{P.}~\bibnamefont{Erd\H{o}s}} \bibnamefont{and}
  \bibinfo{author}{\bibfnamefont{A.}~\bibnamefont{R\'{e}nyi}},
  \bibinfo{journal}{Publicationes mathematicae} \textbf{\bibinfo{volume}{6}},
  \bibinfo{pages}{290} (\bibinfo{year}{1959}).

\bibitem[{\citenamefont{Newman et~al.}(2001)\citenamefont{Newman, Strogatz, and
  Watts}}]{newman-2001random}
\bibinfo{author}{\bibfnamefont{M.~E.} \bibnamefont{Newman}},
  \bibinfo{author}{\bibfnamefont{S.~H.} \bibnamefont{Strogatz}},
  \bibnamefont{and} \bibinfo{author}{\bibfnamefont{D.~J.} \bibnamefont{Watts}},
  \bibinfo{journal}{Physical Review E} \textbf{\bibinfo{volume}{64}},
  \bibinfo{pages}{026118} (\bibinfo{year}{2001}).

\bibitem[{\citenamefont{Albert et~al.}(2000)\citenamefont{Albert, Jeong, and
  Barab{\'a}si}}]{albert2000error}
\bibinfo{author}{\bibfnamefont{R.}~\bibnamefont{Albert}},
  \bibinfo{author}{\bibfnamefont{H.}~\bibnamefont{Jeong}}, \bibnamefont{and}
  \bibinfo{author}{\bibfnamefont{A.-L.} \bibnamefont{Barab{\'a}si}},
  \bibinfo{journal}{Nature} \textbf{\bibinfo{volume}{406}},
  \bibinfo{pages}{378} (\bibinfo{year}{2000}).

\bibitem[{\citenamefont{Blakley}(1979)}]{blakley1899safeguarding}
\bibinfo{author}{\bibfnamefont{G.~R.} \bibnamefont{Blakley}}, in
  \emph{\bibinfo{booktitle}{Managing Requirements Knowledge, International
  Workshop on}} (\bibinfo{organization}{IEEE Computer Society},
  \bibinfo{year}{1979}), pp. \bibinfo{pages}{313--317}.

\bibitem[{\citenamefont{Shamir}(1979)}]{shamir1979share}
\bibinfo{author}{\bibfnamefont{A.}~\bibnamefont{Shamir}},
  \bibinfo{journal}{Communications of the ACM} \textbf{\bibinfo{volume}{22}},
  \bibinfo{pages}{612} (\bibinfo{year}{1979}).

\bibitem[{\citenamefont{Dolev et~al.}(1993)\citenamefont{Dolev, Dwork, Waarts,
  and Yung}}]{dolev-acm1993}
\bibinfo{author}{\bibfnamefont{D.}~\bibnamefont{Dolev}},
  \bibinfo{author}{\bibfnamefont{C.}~\bibnamefont{Dwork}},
  \bibinfo{author}{\bibfnamefont{O.}~\bibnamefont{Waarts}}, \bibnamefont{and}
  \bibinfo{author}{\bibfnamefont{M.}~\bibnamefont{Yung}}, \bibinfo{journal}{J.
  ACM} \textbf{\bibinfo{volume}{40}}, \bibinfo{pages}{17}
  (\bibinfo{year}{1993}), ISSN \bibinfo{issn}{0004-5411},
  \urlprefix\url{http://doi.acm.org/10.1145/138027.138036}.

\bibitem[{\citenamefont{Vitali et~al.}(2011)\citenamefont{Vitali, Glattfelder,
  and Battiston}}]{vitali-plosone2011}
\bibinfo{author}{\bibfnamefont{S.}~\bibnamefont{Vitali}},
  \bibinfo{author}{\bibfnamefont{J.~B.} \bibnamefont{Glattfelder}},
  \bibnamefont{and}
  \bibinfo{author}{\bibfnamefont{S.}~\bibnamefont{Battiston}},
  \bibinfo{journal}{PLoS ONE} \textbf{\bibinfo{volume}{6}},
  \bibinfo{pages}{e25995} (\bibinfo{year}{2011}),
  \urlprefix\url{http://dx.doi.org/10.1371%2Fjournal.pone.0025995}.

\bibitem[{\citenamefont{Battiston et~al.}(2012)\citenamefont{Battiston, Puliga,
  Kaushik, Tasca, and Caldarelli}}]{battiston-sreps2012}
\bibinfo{author}{\bibfnamefont{S.}~\bibnamefont{Battiston}},
  \bibinfo{author}{\bibfnamefont{M.}~\bibnamefont{Puliga}},
  \bibinfo{author}{\bibfnamefont{R.}~\bibnamefont{Kaushik}},
  \bibinfo{author}{\bibfnamefont{P.}~\bibnamefont{Tasca}}, \bibnamefont{and}
  \bibinfo{author}{\bibfnamefont{G.}~\bibnamefont{Caldarelli}},
  \bibinfo{journal}{Sci. Rep.} \textbf{\bibinfo{volume}{2}}
  (\bibinfo{year}{2012}), ISSN \bibinfo{issn}{2045-2322},
  \urlprefix\url{http://dx.doi.org/10.1038/srep00541}.

\bibitem[{\citenamefont{Tessone et~al.}(2013)\citenamefont{Tessone, Garas,
  Guerra, and Schweitzer}}]{tessone-jstatphys2013}
\bibinfo{author}{\bibfnamefont{C.~J.} \bibnamefont{Tessone}},
  \bibinfo{author}{\bibfnamefont{A.}~\bibnamefont{Garas}},
  \bibinfo{author}{\bibfnamefont{B.}~\bibnamefont{Guerra}}, \bibnamefont{and}
  \bibinfo{author}{\bibfnamefont{F.}~\bibnamefont{Schweitzer}},
  \bibinfo{journal}{Journal of Statistical Physics}
  \textbf{\bibinfo{volume}{151}}, \bibinfo{pages}{765} (\bibinfo{year}{2013}),
  ISSN \bibinfo{issn}{0022-4715},
  \urlprefix\url{http://dx.doi.org/10.1007/s10955-013-0723-y}.

\bibitem[{\citenamefont{Masuda and Konno}(2006)}]{masuda-jtheoretbio2006}
\bibinfo{author}{\bibfnamefont{N.}~\bibnamefont{Masuda}} \bibnamefont{and}
  \bibinfo{author}{\bibfnamefont{N.}~\bibnamefont{Konno}},
  \bibinfo{journal}{Journal of Theoretical Biology}
  \textbf{\bibinfo{volume}{243}}, \bibinfo{pages}{64 } (\bibinfo{year}{2006}),
  ISSN \bibinfo{issn}{0022-5193},
  \urlprefix\url{http://www.sciencedirect.com/science/article/pii/S00225193060%
02438}.

\end{thebibliography}

\end{document}